\documentclass[useAMS,usenatbib]{mn2e}
\pdfoutput=1
\usepackage{graphicx}
\usepackage{txfonts}
\usepackage[breaklinks, colorlinks, citecolor=cyan]{hyperref}
\usepackage{url}
\usepackage{natbib}
\usepackage{color}
\usepackage{amssymb}
\usepackage{booktabs}
\usepackage[table]{xcolor}
\definecolor{tableShade}{HTML}{F1F5FA}   
\definecolor{tableShade2}{HTML}{ECF3FE} 
\definecolor{light-grey}{gray}{0.90}

\newfont{\gwpfont}{cmssq8 scaled 1000}
\newcommand{\rexcess}{{\gwpfont REXCESS}}

\newcommand{\msun}{{~{\rm M}_\odot}}

\newcommand{\etal}{{et~al.}~}

\newcommand{\chandra}{{\it{Chandra}}}
\newcommand{\suzaku}{{\it{Suzaku}}}
\newcommand{\mg}{${\rm{M}}_{\rm{g}}$}
\newcommand{\fg}{${\rm{f}}_{\rm{gas}}$}
\newcommand{\tx}{${\rm{T}}_{\rm{X}}$}
\newcommand{\yx}{${\rm{Y}}_{\rm{X}}$}
\newcommand{\lx}{${\rm{L}}_{\rm{X}}$}
\newcommand{\rt}{${\rm{R}}_{2500}$}
\newcommand{\rf}{${\rm{R}}_{500}$}

\title[X-ray Properties of Optically Selected Clusters]{The X-ray Properties of Optically Selected Clusters of Galaxies}

\author[A. K. Hicks et al.]{A. K. Hicks,$^{1,2}$\thanks{E-mail:
ahicks@alum.mit.edu}
G. W. Pratt,$^{3}$
M. Donahue,$^{2}$
E. Ellingson,$^{4}$
M. Gladders,$^{5}$
\newauthor
H. B\"{o}hringer,$^{6}$
H. K. C. Yee,$^{7}$
R. Yan,$^{8}$
J. H. Croston,$^{9}$
and D. G. Gilbank$^{10}$\\
$^{1}$Eureka Scientific, 2452 Delmer Street, Suite 100, Oakland, CA 94602-3017, USA\\
$^{2}$Department of Physics \& Astronomy, Michigan State University, East Lansing, MI 48824-2320, USA\\
$^{3}$Laboratoire AIM, IRFU/Service d'Astrophysique - CEA/DSM - CNRS - Universit\'{e} Paris Diderot, B\^{a}t. 709, \\ \indent CEA-Saclay, 91191 Gif-sur-Yvette Cedex, France\\
$^{4}$Center for Astrophysics and Space Astronomy, University of Colorado at Boulder, Campus Box 389, Boulder, CO 80309, USA\\
$^{5}$Department of Astonomy and Astrophysics, University of Chicago, 5640 S. Ellis Ave, Chicago, IL 60637, USA\\
$^{6}$Max-Planck-Institut fŸr extraterrestriche Physik, Giessenbachstra§e, 85748 Garching, Germany \\
$^{7}$Department of Astronomy and Astrophysics, University of Toronto, 50 St. George St., Toronto, ON, M5S 3H4, Canada\\
$^{8}$Department of Physics and Astronomy, University of Kentucky, 505 Rose St. Lexington, KY, 40506-0055\\
$^{9}$School of Physics and Astronomy, University of Southampton, Southampton, Hampshire, SO17 1BJ, UK\\
$^{10}$South African Astronomical Observatory, PO Box 9, Observatory 7935, South Africa}

\begin{document}

\date{Received 2012 June}


\maketitle

\begin{abstract}
We present the results of \chandra~and \suzaku~X-ray observations of nine moderate-redshift ($0.16 < z < 0.42$) clusters discovered via the Red-sequence Cluster Survey (RCS).
Surface brightness profiles are fitted to beta models, gas masses are determined, integrated spectra are extracted within \rt, and X-ray temperatures and luminosities are inferred. 
The \lx-\tx~relationship expected from self-similar evolution is tested by comparing this sample to our previous X-ray investigation of nine high-redshift ($0.6 < z < 1.0$) optically selected clusters. We find that optically selected clusters are systematically less luminous than X-ray selected clusters of similar X-ray temperature at both moderate and high-$z$. We are unable to constrain evolution in the \lx-\tx ~relation with these data, but find it consistent with no evolution, within relatively large uncertainties.
To investigate selection effects, we compare the X-ray properties of our sample to those of clusters in the representative X-ray selected \rexcess~sample, also determined within \rt.  We find that while RCS cluster X-ray properties span the entire range of those of massive clusters selected by other methods, their average X-ray properties are most similar to those of dynamically disturbed X-ray selected clusters.  This similarity suggests that the true cluster distribution might contain a higher fraction of disturbed objects than are typically detected in X-ray selected surveys.

\end{abstract}


\section{Introduction \label{s:intro}}

By virtue of their size, clusters of galaxies are an important source of information about the underlying cosmology of the universe~\citep[e.g.,][and references therein]{allen11}. Since the advent of the first cluster survey over 50 years ago~\citep{abell58}, the search for these impressive objects has grown in cosmological impact, and is currently being pursued with the ultimate objective of constraining $w$, the dark energy equation of state. Improved technology has enabled large-area cluster searches in several wavebands (e.g., X-ray, optical, millimetre, submm, radio) and out to formative redshifts. Consequently, the number of recent, underway, and planned cluster surveys is staggering: \citealt{bohringer04}~\citetext{REFLEX};~\citealt{valtchanov04}~\citetext{XMM-LSS};~\citealt{gladders05}~\citetext{RCS};~\citealt{wester05}~\citetext{DES};~\citealt{content08}~\citetext{EUCLID};~\citealt{wilson09}~\citetext{SpARCS};~\citealt{conconi10}~\citetext{WFXT};~\citealt{vanderlinde10}~\citetext{SPT};~\citealt[][]{Lloyd-Davies11}~\citetext{XCS};~\citealt{marriage11}~\citetext{ACT};~\citealt{planck2011-5.1a}~\citetext{Planck};~\citealt{predehl11}~\citetext{eROSITA};~\citealt{suhada12}~\citetext{XMM-BCS} and many more.

\begin{table*}
\centering
\begin{minipage}{120mm} 
\caption{Cluster Sample\label{table1}}
\begin{tabular}{ccccccc}
\hline
\multicolumn{2}{c}{Cluster} & 
{{\rm{z}}} & 
{$1\arcsec$} &
{Telescope} &
{obsid} &
{Exposure}\\
\multicolumn{2}{c}{} & 
 &
{[$h_{70}^{-1}$ kpc]} &
 &
& 
{[seconds]} \\
\hline
\multicolumn{2}{l}{RCS0222+0144} & 0.25\footnote[1]{Spectroscopic~\citep{blindert07}} & 3.91  & {\it{Chandra}}  &10485  & 23,335 \\
\multicolumn{2}{l}{RCS1102-0319} & 0.33\footnote[2]{From X-ray spectra (this work), see text.} &  4.75 &{\it{Suzaku}}   &  803065010& 26,381\\
\multicolumn{2}{l}{RCS1102-0340} & 0.39\footnote[3]{Spectroscopic~\citep{ellingson12}} & 5.29  &{\it{Suzaku}}   &  803064010& 37,797\\
\multicolumn{2}{l}{RCS1330+3043} & 0.27\footnote[4]{Photometric} &  4.14 & {\it{Chandra}}  & 10487 & 15,824\\
\multicolumn{2}{l}{RCS1447+0828} & $0.38^{\displaystyle c}$ & 5.21  & {\it{Chandra}}  &10481  &11,999 \\
\multicolumn{2}{l}{RCS1447+0949} & $0.20^{\displaystyle a}$ & 3.30  &{\it{Chandra}}   &  10486& 17,540\\
\multicolumn{2}{l}{RCS1615+3057} & $0.42^{\displaystyle a}$ & 5.53  &{\it{Chandra}}   &  10482& 27,777\\
\multicolumn{2}{l}{RCS2150-0442} & $0.16^{\displaystyle b}$ &  2.76 & {\it{Chandra}}  &  10488 & 10,506 \\
\multicolumn{2}{l}{RCS2347-3535} & $0.26^{\displaystyle a}$ &  4.02 & {\it{Chandra/Suzaku}}  & 10484/803057010 & 13,850/15,489 \\
\hline
\end{tabular}
\end{minipage}
\end{table*}

One overarching goal of these endeavors is to chart the evolution of the cluster mass function, thereby providing key constraints on the progression of large-scale structure formation in the universe. Constructing a mass function requires two key elements: the ability to find clusters, and an efficient method of mass estimation. Obtaining either piece necessitates an accurate understanding of relationships between the observable properties of clusters (i.e., baryons) and their underlying dark matter distributions.  

The baryonic mass in clusters is dominated by the hot intracluster medium (ICM), therefore some of the most powerful methods of cluster identification (X-ray, SZ) use the ICM to both find clusters and estimate their masses.  Several processes occur within clusters, however, that can alter the physical characteristics and distribution of cluster baryons.  These include feedback~\citep[e.g.,][]{voit05}, mergers~\citep[e.g.,][]{chatzikos12}, and radiative cooling~\citep[e.g.,][]{pratt09}, among other non-gravitational processes~\citep[e.g.,][]{nagai06}.  Additionally, the frequency with which those processes occur may vary with redshift~\citep[e.g.,][]{barger05}. 

Understanding the gamut of ICM properties requires X-ray observations of samples chosen independently of their X-ray characteristics. This has indeed been pursued multiple times~\citep[e.g.,][]{holden97,donahue01,donahue02,basilakos04,gilbank04,popesso04,sanchez05}. Most of these surveys have identified a significant population of low-\lx~clusters. Unfortunately, there is a striking dearth of adequate X-ray data for these samples. Without convincing \lx~and \tx~measurements it is impossible to verify alternatively obtained cluster masses or investigate trends in core gas density. Due largely to this lack of quality X-ray data, concrete physical explanations for the observed scatter in cluster properties have not yet been well determined.

In our previous work~\citep{hicks08} we use pointed \chandra~observations to compare the ICM properties of high-$z$ optically selected clusters~\citep[RCS;][]{gladders05} with those of moderate-redshift X-ray selected clusters~(CNOC; e.g.,~\citealt[][]{yee96b}).  Our results indicate that the typical central ($\sim$\rt; where $\rho_{clust}/\rho_{crit}=2500$) gas mass fractions of the RCS sample are strikingly lower than those found in X-ray selected systems.  This is also illustrated by discrepancies in the normalization of the \lx-\tx~relationship between the two samples.  These comparisons, however, were made between clusters of differing average redshift ($z_{\rm{CNOC}} \sim 0.3$, $z_{\rm{RCS}}\sim0.8$), which were gathered via significantly different means of selection, making it difficult to disentangle the effects of selection bias from possible signatures of cluster evolution.

The main objectives of this work are to isolate these potential factors from one another and identify the cause of discrepancies in ICM properties between optically and X-ray selected cluster samples.  In Section~2 we introduce the present cluster sample and describe the initial processing of our \chandra~and \suzaku~X-ray observations. In Sections~3 and~4 we probe the surface brightness, temperature, metallicity, and density of the hot ICM present in each cluster. In Section~5 we compare the ICM properties of RCS clusters to X-ray selected samples, 
including the \rexcess~sample for which quantities have been extracted within \rt.  Finally, we compare X-ray temperature to velocity dispersion for a subset of our targets (Section~6).  Our results are summarized in Section~7. 

Unless otherwise noted, this paper assumes a cosmology of $\Omega_{\rm{M}}=0.3$, $\Omega_{\Lambda}=0.7$ and $\rm{H}_0=70~\rm{km}~\rm{s}^{-1}~\rm{Mpc}^{-1}$.  All errors are quoted at 68\% confidence levels.

\section{Cluster Sample \& Observations \label{s:obs}}  

In an attempt to decouple the effects of sample selection from possible redshift evolution in X-ray properties, our targets were chosen to match our X-ray selected CNOC comparison sample in both redshift and mass.  Velocity dispersion was used as a mass proxy for two-thirds of our targets, and initial masses for the remaining third were estimated from cluster richness~\citep{yee99,gladders05}.  The ranges of redshifts and velocity dispersions spanned by our sample are $0.16<z<0.42$ and $709<\sigma<1390$ km s$^{-1}$, compared to CNOC ranges of $0.17<z<0.55$ and $575<\sigma<1330 $ km s$^{-1}$.  Each target was observed in the X-ray by either {\it{Chandra}} or {\it{Suzaku}}.  Table~\ref{table1} lists the clusters in our sample along with their redshifts, the telescope used, obsids, and exposure times.

\begin{figure*}
\centering
\includegraphics[width=6in]{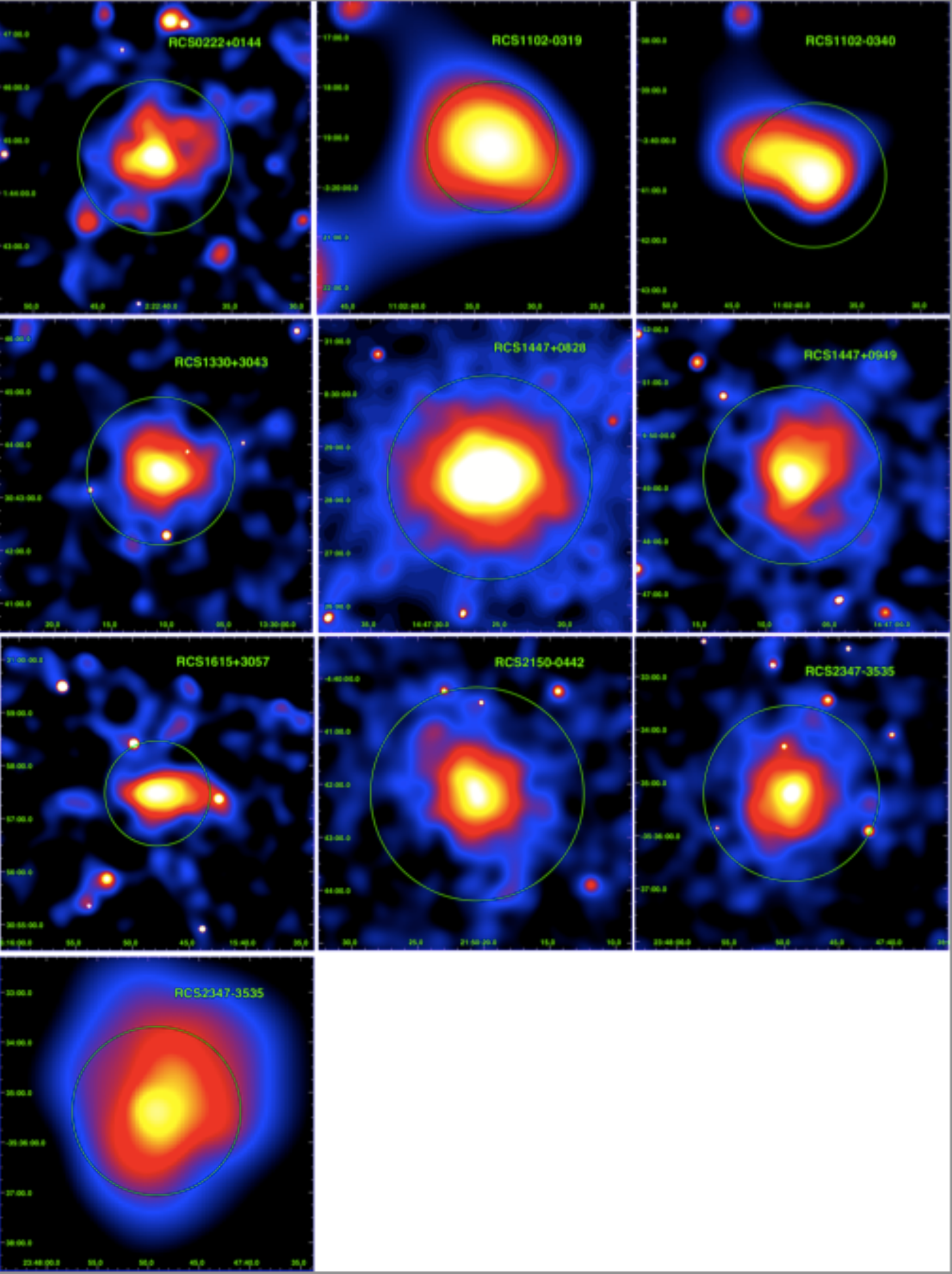}
\caption{{\bf{Smoothed Flux Images.}}  Adaptively smoothed 0.3-7.0 keV \chandra~flux and 0.2-12.0 keV \suzaku~counts images of our sample.  RCS1102-0319, RCS1102-0340, and the second RCS2347-3535 panel are \suzaku~observations.  Circles denote calculated values of $\rm{R}_{2500}$ for each cluster.  In each image, north is up and east is to the left.  
\label{fig1}}
\end{figure*}
\begin{table*}
\centering
\begin{minipage}{150mm} 
\caption{Cluster Positions and Detection Details\label{table2}}
\begin{tabular}{ccllllccc}
\hline
\multicolumn{2}{c}{Cluster} & 
\multicolumn{2}{c}{Optical Position\footnote[1]{All positions are given for equinox J2000}} &
\multicolumn{2}{c}{X-ray Peak\footnote[2]{0.3-7.0 keV band, within ${\rm{R}} < 500 ~h^{-1}_{70}~{\rm{kpc}}$}} &
{Separation} &
{Net Counts$^{\displaystyle b}$} &
{S/N Ratio}   \\
\multicolumn{2}{c}{} & 
{RA} &
{Dec} &
{RA} &
{Dec} &
{[$\arcsec$]}&
{} &
{} 
\\
\hline
\multicolumn{2}{l}{RCS0222+0144} & 02 22 40.6 & +01 44 32.0 & 02 22 40.8 & +01 44 40.3 & 8.8 & 1036 & 14.8 \\
\multicolumn{2}{l}{RCS1102-0319} & 11 02 33.0 & -03 19 04.8 & 11 02 33.6 & -03 19 12.2  & 11.6 &  2440\footnote[3]{{\it{Suzaku}} observation, $\rm{R}=260$\arcsec}& 29.9$^{\displaystyle c}$\\
\multicolumn{2}{l}{RCS1102-0340} & 11 02 39.1 & -03 40 17.1 & 11 02 38.6 & -03 40 42.7 & 26.7 &  4803$^{\displaystyle c}$ & 43.8$^{\displaystyle c}$\\
\multicolumn{2}{l}{RCS1330+3043} & 13 30 10.7 & +30 43 30.6 & 13 30 10.7 & +30 43 30.0 & 0.6 & 926 & 16.2\\
\multicolumn{2}{l}{RCS1447+0828} & 14 47 26.9 & +08 28 17.5 & 14 47 25.9 & +08 28 25.1 & 16.7 & 14022 & 112.8\\
\multicolumn{2}{l}{RCS1447+0949} & 14 47 08.1 & +09 49 01.7 & 14 47 08.0 & +09 49 14.2 & 12.6 & 1531 & 18.7\\
\multicolumn{2}{l}{RCS1615+3057} & 16 15 47.2 & +30 57 18.0 & 16 15 47.8 & +30 57 28.4 & 13.0 & 339 & 11.1 \\
\multicolumn{2}{l}{RCS2150-0442} & 21 50 19.7 & -04 42 25.4 & 21 50 20.5 &  -04 42 11.2 & 18.6 & 3001 & 48.3\\
\multicolumn{2}{l}{RCS2347-3535} & 23 47 49.2 & -35 35 10.9 & 23 47 49.4\footnote[4]{Using {\it{Chandra}} data} & -35 35 12.3$^{\displaystyle d}$ & 2.8 &2911  &54.6 \\
\multicolumn{2}{l}{}&  &  & 23 47 49.2\footnote[5]{Using {\it{Suzaku}} data} & -35 35 22.5$^{\displaystyle e}$ & 11.6 &  3823$^{\displaystyle c}$& 60.2$^{\displaystyle c}$\\
\hline
\end{tabular}
\end{minipage}
\end{table*}

\subsection{{\it{Chandra}} Initial Processing}

Seven of our nine targets were observed with {\it{Chandra's}} ACIS-S (Advanced CCD Imaging Spectrometer) array in VFAINT mode during the period 2009 November 24 - 2010 August 3.  The field of view of the ACIS-S aimpoint chip is 8'x8'.  \chandra~has exceptional 0.5\arcsec spatial resolution, enabling very detailed investigations of cluster cores.  The spectral resolution of ACIS is 120-130 eV over the energy range of interest to this study (0.3-7 keV).

Preliminary reduction of the~\chandra~data was performed as in~\citet{hicks08}.  
After this initial cleaning, 0.3-7.0 keV and 0.3-2.5 keV images, instrument maps, and exposure maps were created  for each dataset using the CIAO 4.2 tool MERGE\textunderscore ALL and CALDB version 4.3.0.  Point source detection was performed by running the tools WTRANSFORM and WRECON on the 0.3-7.0 keV flux images.  Data with energies below 0.3 keV and above 7.0 keV were excluded due to uncertainties in the ACIS calibration and background contamination, respectively. These products were used solely for imaging analysis; our spectral analysis excludes all data below 0.6 keV.

\subsection{{\it{Suzaku}} Initial Processing}

The {\it{Suzaku}} X-ray Imaging Spectrometer (XIS) observed three of the clusters in our sample between May 31 and December 17, 2008 including one that was later observed by {\it{Chandra}} (RCS2347-3535).  Suzaku's field of view is 18\arcmin x18\arcmin, and it has a spatial resolution of only $\sim2$\arcmin.  One of the advantages of~\suzaku, however, is its excellent low-energy spectral resolution (60 eV at 0.25 keV).

\suzaku~data were reprocessed to include the most recent calibration files using the HEASOFT 6.9 {\it{Suzaku}} software Version 16 tool AEPIPELINE.  Calibration sources were removed from the data using XSELECT, and data from XIS0, XIS1, and XIS3 were combined to produce integrated images in the 0.2-12.0 keV energy range. 

\section{Images and surface Brightness profile modelling\label{s:sb}}

\subsection{Images and cluster centers}

Figure~\ref{fig1} contains smoothed 0.3-7.0 keV {\it{Chandra}} flux and 0.2-12.0 keV {\it{Suzaku}} counts images of each of the clusters in our sample (produced by the CIAO tool CSMOOTH).  Using these images we determined the location of the X-ray emission peak of each cluster.  Optical positions were taken from RCS cluster catalogs, which were computed from the smoothed galaxy distribution used for cluster identification~\citep[see][for details]{gladders05}.  \suzaku~observations were included in this exercise despite having $\sim2\arcmin$ spatial resolution.
All clusters were found within $27\arcsec$ of their optical positions, and all \chandra~observed clusters were found within 19$\arcsec$, with a mean offset of 10.4$\arcsec$.  
Table~\ref{table2} lists optical positions, X-ray positions, net counts ($C-B$) and signal-to-noise ratios derived from the method described in Appendix~\ref{s:sig}. All clusters are detected at high significance. The median number of source counts per observation is 2911, with values ranging between 339 (RCS1615+3057) and 14,022 (RCS1447+0828).

\subsection{Surface Brightness Profiles}\label{sec:beta}

Radial surface brightness profiles were extracted in circular annuli from both targeted observations and blank sky background files using 0.3-2.5 keV \chandra~counts images. The cluster profiles were binned to ensure S/N $>3$ in each bin and background-subtracted.  None of our surface brightness bins are narrower than 1\arcsec, therefore we do not expect to see any effects from~\chandra's~0.5\arcsec~FWHM PSF on our profiles.

The surface brightness profiles were fitted to a beta-model:

\begin{equation}
I(r) = I_0 \left( 1 + {r^2 \over r_c^2} \right)^{-3\beta+\frac{1}{2}},
\label{sb_eq}
\end{equation} 

\noindent where $I_{0}$ is the normalization and $\rm{r}_{\rm{c}}$ is the core radius. We compared the measured surface brightness (with error bars computed from photon statistics) to the average model surface brightness integrated within each radial bin while performing a Levenberg-Marquardt least-squares fit to the model, as implemented in the IDL routine mpcurvefit{\footnote{Available courtesy Craig B. Markwardt, http://cow.physics.wisc.edu/~craigm/idl/idl.html}}.

Best fitting model parameters for the seven \chandra-observed clusters are given in Table \ref{table3}, and images of these fits are shown in Figure~\ref{figB1}. Though many of the clusters exhibit some substructure, most were reasonably well fitted by the $\beta$ model (see Table~\ref{table3} for goodness of fit data).  $\beta$ values for our sample are generally lower than the typical value of 2/3~\citep{neumann99}, with a sample average of only $0.42\pm{0.01}$, suggesting that the gas is less centrally-concentrated in these targets compared to those studied by~\citet{neumann99}.

\begin{table}
\centering
\caption{$\beta$-Model Fits\label{table3}}
\resizebox{\columnwidth}{!} {
\begin{tabular}{ccrrrcr}
\hline
\multicolumn{2}{c}{Cluster} & {$r_{\rm c}$} [kpc] &
{$\beta$}  & {$I_{\rm{0}}$}\footnote[1]{{Surface brightness $I$ in units of $10^{-6}$ photons sec$^{-1}$ pixel$^{-2}$}} & Outermost Bin [kpc]\footnote[2]{Radial range of outermost bin used in fitting procedure, in kpc.} &
{$\chi^2/\rm{DOF}$}\\
\hline
 \multicolumn{2}{l}{RCS0222+0144}& $24\pm{13}$ & $0.32\pm{0.02}$ & $2.0\pm{0.7}$ & 276-297 &11.0/11 \\ 
 \multicolumn{2}{l}{RCS1330+3043}& $30\pm{8}$ & $0.39\pm{0.02}$ & $5\pm{1}$ & 286-366  &28.3/28  \\ 
 \multicolumn{2}{l}{RCS1447+0828}& $28.9\pm{0.8}$ & $0.581\pm{0.003}$ & $820\pm{20}$ & 836-875 &135.6/116 \\ 
 \multicolumn{2}{l}{RCS1447+0949}& $30\pm{7}$ & $0.36\pm{0.01}$ & $4.4\pm{0.7}$ & 450-515 &40.6/52 \\ 
 \multicolumn{2}{l}{RCS1615+3057}& $37\pm{22}$ & $0.34\pm{0.04}$ & $1.8\pm{0.6}$ &235-276  &6.5/8  \\ 
 \multicolumn{2}{l}{RCS2150-0442}& $55\pm{6}$ & $0.51\pm{0.02}$ & $9.1\pm{0.8}$ & 370-411 &63.8/64  \\ 
  \multicolumn{2}{l}{RCS2347-3535}& $70\pm{12}$ & $0.45\pm{0.02}$ & $5.1\pm{0.6}$  & 454-476 &53.2/53  \\ 
\hline
\end{tabular}
}
\end{table}

RCS2347-3535 contains a region of excess emission in its central 10-20 kpc, suggesting that it may harbour a small cool core.  None of the other clusters in our sample exhibit a significant central surface brightness excess above the $\beta$ model.

\begin{table*}
\begin{center}
\begin{minipage}{140mm}
\caption{Spectral Fits ($\Delta = 2500$)\label{table4}}
\begin{tabular}{cccccccc}
\hline
\multicolumn{2}{c}{Cluster} & {$z$} & ${\rm{N}}_{\rm{H}}$ & {kT} & {Z} & {Norm} & {n$_{\rm{e,0}}$}\\
\multicolumn{2}{c}{} 
& {} 
& {[${10}^{20}~{\rm{cm}}^{-2}$]} 
& {[keV]} 
& {[Z$_{\sun}$]}  
& {[${10}^{-4}~{\rm{cm}}^{-5}$]}
& {[$10^{-3}$~cm$^{-3}$]} \\
\hline
\multicolumn{2}{l}{RCS0222+0144} & 0.25  & 3.30 & {${4.6}^{+0.9}_{-0.8}$} & {${1.4}^{+0.9}_{-0.6}$}   & {${1.7}^{+0.3}_{-0.3}$} & {${4.4}^{+0.3}_{-0.4}$}\\
\multicolumn{2}{l}{RCS1102-0319\footnote[1]{{\it{Suzaku}} observation, $\rm{R}=260$\arcsec}} &  $0.334^{+0.007}_{-0.007}$ & 4.04 & {${4.3}^{+0.5}_{-0.4}$} & {${0.5}^{+0.2}_{-0.2}$}  &   {${3.7}^{+0.3}_{-0.3}$}  & \ldots \\
\multicolumn{2}{l}{RCS1102-0340$^{\displaystyle a}$} &  0.39 &4.06 & {${6.3}^{+0.7}_{-0.7}$} &  ${0.2}^{+0.1}_{-0.1}$   &   {${5.8}^{+0.2}_{-0.2}$} & \ldots \\
\multicolumn{2}{l}{RCS1330+3043} & 0.27 &1.16 & {${4.4}^{+1.0}_{-0.9}$} &  ${0.3}^{+0.3}_{-0.3}$   &  {${3.6}^{+0.5}_{-0.3}$}& {${8.9}^{+0.6}_{-0.4}$}\\
\multicolumn{2}{l}{RCS1447+0828}&  0.38 &2.24 & {${6.8}^{+0.3}_{-0.2}$} &  ${0.38}^{+0.06}_{-0.05}$   & {${101}^{+1}_{-1}$}  & {${161}^{+1}_{-1}$} \\ 
\multicolumn{2}{r}{(Core-excised)\footnote[2]{Central $0.15~\rm{R}_{500}$ removed from spectrum.}} & \ldots & 2.24 & {${13}^{+3}_{-2}$} &  ${0.4}^{+0.3}_{-0.2}$   & {${24}^{+1}_{-1}$}  & \ldots \\
\multicolumn{2}{l}{RCS1447+0949}&  0.20 &1.83& {${2.4}^{+0.3}_{-0.3}$} &  ${0.02}^{+0.09}_{-0.02}$   &  {${6.8}^{+0.5}_{-0.5}$}  & {${7.7}^{+0.3}_{-0.3}$} \\
\multicolumn{2}{l}{RCS1615+3057} &  0.42 &2.6 & {${3.8}^{+1}_{-0.9}$} &   ${0.0}^{+0.1}_{-0.0}$   &  {${1.7}^{+0.1}_{-0.2}$}  & {${5.5}^{+0.2}_{-0.3}$} \\
\multicolumn{2}{l}{RCS2150-0442} &  $0.160^{+0.008}_{-0.015}$ & 2.72& {${3.6}^{+0.4}_{-0.3}$} &  ${0.4}^{+0.2}_{-0.1}$   &   {${11.6}^{+0.6}_{-0.5}$} & {${8.2}^{+0.2}_{-0.3}$} \\
\multicolumn{2}{l}{RCS2347-3535\footnote[3]{Best simultaneous fit to {\it{Chandra}} and {\it{Suzaku}} spectra.}} &  0.26 &1.22& {${4.9}^{+0.3}_{-0.3}$} &  ${0.49}^{+0.10}_{-0.09}$  & ${7.3}^{+0.3}_{-0.3}$\footnote[4]{{\it{Chandra}} normalization; {\it{Suzaku}} normalization for this fit is ${11.5}^{+0.4}_{-0.4}\times10^{-4}~{\rm{cm}}^{-5}$.} & {${5.9}^{+0.1}_{-0.1}$}\\
\multicolumn{2}{l}{RCS2347-3535\footnote[5]{Using only {\it{Chandra}} data.}} & \ldots  & 1.22& {${5.3}^{+0.7}_{-0.7}$} &  ${0.5}^{+0.3}_{-0.2}$   &  {${7.3}^{+0.4}_{-0.5}$}   & {${5.9}^{+0.1}_{-0.1}$}\\
\multicolumn{2}{r}{(Core-excised)$^{\displaystyle b,e}$} & \ldots  & 1.22& {${4.9}^{+1.3}_{-0.8}$} &  ${0.08}^{+0.32}_{-0.08}$   &  {${4.7}^{+0.3}_{-0.4}$}  & \ldots \\
\multicolumn{2}{l}{RCS2347-3535$^{\displaystyle a}$\footnote[6]{Using only {\it{Suzaku}} data.}} & \ldots  &1.22& {${4.7}^{+0.3}_{-0.3}$} &  ${0.5}^{+0.1}_{-0.1}$   &  {${11.5}^{+0.4}_{-0.4}$} & \ldots \\
\hline
\end{tabular}
\end{minipage}
\end{center}
\end{table*}

\section{Spectroscopy\label{s:integrated}}

\subsection{Initial spectral fits and redshift estimates}\label{sec:init}

Initial spectra were extracted from each point-source-removed event file in a circular region with a $300~\rm{h}_{70}^{-1}~\rm{kpc}$ radius (\chandra) or 260\arcsec~radius (\suzaku).
These values were chosen to provide an initial spectrum with as close to an~\rt~extraction radius as possible.  In the case of \suzaku~observations, the smallest recommended radius for extracting spectral data is 260\arcsec.  Individual spectra were extracted from each \suzaku~XIS CCD and fitted simultaneously.  

All spectra were analyzed in XSPEC~\citep{arnaud96}, using weighted response matrices (RMFs) and effective area files (ARFs) generated with the CIAO tool SPECEXTRACT (\chandra), and XISRMFGEN and XISSIMARFGEN (\suzaku).  For the \chandra~observations, blank sky backgrounds were renormalized to match the 9.5-12.0 keV count rates of each exposure, then extracted in identical regions as source spectra.  \suzaku~backgrounds were compiled from the remaining data after calibration sources, point sources, and the target cluster were removed.

Spectra were fitted with single temperature spectral models, inclusive of
foreground absorption, using Cash statistics~\citep{cash79}.  
Each spectrum was fitted with the absorbing column frozen at its measured value~\citep{dickey90}. Metal abundances were allowed to vary.  Data with energies below 0.6 keV and above 7.0 keV were excluded from the \chandra~fits, while \suzaku~spectra were fitted in the 0.5-8.0 keV energy range~\citep[e.g.,][]{bautz07}.  

Redshifts were fitted for the three clusters in our sample that do not have spectroscopic measurements.  We were unable to constrain a redshift for RCS1330+3043, and therefore use its photometric redshift ($z=0.27$) in further analysis.  The fits of the other two clusters resulted in $z=0.334\pm{0.007}$ for RCS1102-0319, and $z=0.160^{+0.008}_{-0.015}$ for RCS2150-0442, within 7\% and 20\% (respectively) of photometric redshift estimates obtained using the color of the red sequence~\citep[e.g.,][]{gladders05}.  
Fits with redshift fixed at these values were used in subsequent analysis (Table~\ref{table4}). The small uncertainties in these values ($\leq10\%$) do not substantially affect our analysis.  

A core-removed (0.15~\rf) spectrum was extracted and fitted for RCS1447+0828, indicating that it possesses  a strong cool core, with an integrated temperature of $6.8^{+0.3}_{-0.2}$ keV and core-removed temperature \tx=$13^{+3}_{-2}$ keV (Table~\ref{table4}). Due to the central excess in its surface brightness profile, a core-excised spectrum was also fitted for RCS2347-3535; however, no conclusive evidence for soft central emission was found in this case.

%
\begin{table*}
\begin{minipage}{\textwidth}
\begin{centering}
\caption{Sample quantities estimated within \rt.}\label{table5}
\newcolumntype{L}{>{\columncolor{white}[0pt][\tabcolsep]}l}
\newcolumntype{R}{>{\columncolor{white}[\tabcolsep][0pt]}l}

\resizebox{\textwidth}{!} {
\begin{tabular}{@{}lrrrrrrrrrrR@{}}
\toprule

\multicolumn{1}{l}{}  & \multicolumn{5}{c}{$M$--$Y_{\rm X}$} & \multicolumn{5}{c}{$M$--$T_{\rm X}$} \\

\cmidrule[0.5pt](lr){2-6}
\cmidrule[0.5pt](lr){7-11}

\multicolumn{1}{l}{Cluster} & \multicolumn{1}{c}{$R_{2500}$} & \multicolumn{1}{c}{$\rm{M}_{\rm{tot}}$} & \multicolumn{1}{c }{\mg} & \multicolumn{1}{c}{\fg} & \multicolumn{1}{c }{\lx} & \multicolumn{1}{l}{$R_{2500}$} & \multicolumn{1}{c}{$\rm{M}_{\rm{tot}}$} & \multicolumn{1}{c }{\mg} & \multicolumn{1}{c}{\fg} & \multicolumn{1}{c }{\lx} \\

\multicolumn{1}{l}{} & \multicolumn{1}{c}{[kpc]} & \multicolumn{1}{c}{[$10^{14}~\msun$]} & \multicolumn{1}{c }{[$10^{13}~ \msun$]} & \multicolumn{1}{c}{} & \multicolumn{1}{c }{[$10^{44}~{\rm{erg}}~{\rm{s}}^{-1}$]} & \multicolumn{1}{l}{[kpc]} & \multicolumn{1}{c}{[$10^{14}~ \msun$]} & \multicolumn{1}{c }{[$10^{13}~\msun$]} & \multicolumn{1}{c}{} & \multicolumn{1}{c }{[$10^{44}~{\rm{erg}}~{\rm{s}}^{-1}$]} \\

\midrule


\noalign{\smallskip}

RCS0222+0144 & $ {346}^{+14}_{-13}$ & ${0.75}^{+0.09}_{-0.09} $&  ${0.25}^{+0.02}_{-0.02} $ & ${0.032}^{+0.005}_{-0.005} $ & ${0.5}^{+0.2}_{-0.2} $ & $ {423}^{+42}_{-37}$ & ${1.4}^{+0.4}_{-0.4} $  & ${0.38}^{+0.03}_{-0.03} $ & ${0.027}^{+0.008}_{-0.008} $ & ${0.5}^{+0.4}_{-0.1} $ \\

RCS1102-0319\footnote[1]{ {\it{Suzaku}} observation, $L_{\rm X,2500}$ estimated within aperture of radius $\rm{R}=260$\arcsec.} & \ldots & \ldots & \ldots & \ldots & \ldots & $ {387}^{+27}_{-23}$ & ${1.2}^{+0.3}_{-0.2} $  & \ldots & \ldots & ${1.7}^{+0.5}_{-0.4}$\\

RCS1102-0340$^{\displaystyle a}$& \ldots & \ldots & \ldots & \ldots & \ldots & $ {457}^{+26}_{-26}$ & ${2.1}^{+0.4}_{-0.4} $  & \ldots & \ldots & ${3.4}^{+0.5}_{-0.3}$ \\

RCS1330+3043 & $ {370}^{+17}_{-15}$ & ${1.0}^{+0.1}_{-0.1} $&  ${0.48}^{+0.03}_{-0.02} $ & ${0.049}^{+0.007}_{-0.006} $ & ${1.1}^{+0.4}_{-0.3} $ 
&  $ {405}^{+46}_{-42}$ & ${1.3}^{+0.4}_{-0.4} $ & ${0.57}^{+0.04}_{-0.03} $ & ${0.04}^{+0.01}_{-0.01} $ & ${1.2}^{+0.4}_{-0.3} $\\

RCS1447+0828\footnote[2]{Core-excised temperature used to determine $R_\Delta$.} & $ {611}^{+29}_{-22}$ & ${4.8}^{+0.7}_{-0.5} $&  ${5.00}^{+0.04}_{-0.04} $ & ${0.10}^{+0.01}_{-0.01} $ &  ${61}^{+3}_{-3} $ &  $ {661}^{+79}_{-55}$ & ${6}^{+2}_{-2} $& ${5.6}^{+0.9}_{-1.3}$  &  ${0.09}^{+0.03}_{-0.03} $  & ${62}^{+4}_{-4} $\\

RCS1447+0949 & $ {338}^{+10}_{-10}$ & ${0.67}^{+0.06}_{-0.06} $  & ${0.40}^{+0.02}_{-0.02} $ & ${0.060}^{+0.006}_{-0.006} $  & ${0.7}^{+0.2}_{-0.1} $ &  $ {314}^{+21}_{-21}$ & ${0.5}^{+0.1}_{-0.1} $& ${0.35}^{+0.01}_{-0.01} $ & ${0.07}^{+0.01}_{-0.01} $ &  ${0.7}^{+0.1}_{-0.2} $ \\

RCS1615+3057 & $ {334}^{+17}_{-16}$ & ${0.8}^{+0.1}_{-0.1} $& ${0.39}^{+0.01}_{-0.02} $ & ${0.047}^{+0.007}_{-0.007} $  & ${0.9}^{+0.2}_{-0.2} $ &  $ {349}^{+46}_{-42}$ & ${0.9}^{+0.4}_{-0.3} $ & ${0.42}^{+0.01}_{-0.02} $ & ${0.05}^{+0.02}_{-0.02} $ & ${0.8}^{+0.4}_{-0.1} $ \\

RCS2150-0442 & $ {361}^{+9}_{-8}$ & ${0.79}^{+0.06}_{-0.05} $& ${0.48}^{+0.01}_{-0.02} $ & ${0.061}^{+0.005}_{-0.005} $ & ${1.4}^{+0.2}_{-0.2} $ &  $ {393}^{+22}_{-17}$ & ${1.0}^{+0.2}_{-0.1} $& ${0.80}^{+0.07}_{-0.09}$  & ${0.055}^{+0.01}_{-0.006} $ &  ${1.4}^{+0.3}_{-0.2} $\\

RCS2347-3535 & $ {406}^{+7}_{-7}$ & ${1.24}^{+0.06}_{-0.06} $&  ${0.73}^{+0.01}_{-0.02} $ & ${0.059}^{+0.003}_{-0.003} $  & ${2.0}^{+0.4}_{-0.4} $ &  $ {434}^{+14}_{-14}$ & ${1.5}^{+0.1}_{-0.1} $  & ${0.82}^{+0.02}_{-0.02} $ & ${0.055}^{+0.004}_{-0.004} $ & ${2.1}^{+0.4}_{-0.4} $\\

\bottomrule
\end{tabular}
}

\end{centering}
\end{minipage}
\end{table*}


\subsection{Scaled aperture estimate}\label{sec:scaling}

With an average of only 3000 counts per target, we do not have sufficient signal to undertake a full hydrostatic mass analysis, or even fit core-excised spectra out to \rf. Due to the shallow nature of our observations, we choose \rt~as the fiducial radius for our measurements.

We employ scaling relations in order to estimate \rt. The present sample contains objects in a variety of dynamical states and thus it is essential to use an appropriate scaling relation. We measure \rt\ for our \chandra\ observations using \yx ($\equiv{\rm{M}}_{\rm{g}}{\rm{T}}_{\rm{X}}$) as a mass proxy. The decision to use \yx\ was motivated by two principal factors. Firstly, the simulations of \citet{kravtsov06} indicate that \yx\ is a robust mass proxy even in the presence of significant dynamical activity. Secondly, use of \yx\ facilitates comparison of our  optically selected sample to the X-ray selected \rexcess~sample, since \rexcess~uses \yx~exclusively. 

We used an iterative process to estimate the cluster masses. First we calculated central densities using the $\beta$ model parameters from Sect.~\ref{sec:beta} and initial spectral fits described in Sect.~\ref{sec:init}:
\begin{equation}
n_{\rm{e,0}}^2={{4\pi d_{ang}^2~(1+z)^2~K~10^{14}}\over{0.82~4 \pi r_c^3~EI}}~{\rm{cm}}^{-6}
\end{equation}
\citep{ettori03}.  Here K is the normalization of the XSPEC model and $EI$ is the emission integral, estimated by integrating the (spherical) emission from the source out to 10 Mpc \citep[see][for details]{ettori03}.  Values of $n_{\rm{e,0}}$ were then used to determine total gas mass by integrating the cluster density profile out to \rf:
\begin{equation}
\rho_{gas}(r) = \rho_0 \left[1 + {{r^2}\over{r_c^2}}\right]^{-3\beta/2},
\label{eq:dens_eq}
\end{equation}
where $\rho_0 = n_{\rm{e,0}}~\mu_e~m_p$, and $\mu_e=1.17$. We then use the M--\yx\ relation of \citet{arnaud07} to estimate the mass and radius of each object. The~\citet{arnaud07}  M-Yx relation uses T[0.15-0.75 \rf] as well as \mg[\rf]. We translate our T[\rt] measurements to T[0.15-0.75~\rf] using the mean T[\rt]/T[0.15-0.75~\rf] ratio from the \rexcess~sample. \rf~was then converted to \rt~using the simple relationship \rt$=0.44$~\rf~\citep{arnaud05}\footnote{Use of a constant \rt/\rf\ ratio assumes that a constant concentration. This is a fair assumption given the mass range of interest \citep[see e.g.,][]{poi05}.}. The process of extracting spectra, fitting temperatures, determining gas mass, and re-estimating the mass and radius \rt\ was repeated until the radius values converged. 

Given the lower spatial resolution of \suzaku, in these cases we instead determined $\rm{R}_{2500}$ using the M-\tx~relationship of~\citet{arnaud05}. The cluster RCS2347-3535 has been observed by both \chandra~and \suzaku. Here \chandra~spatial information was used in tandem with combined temperature fitting of both \chandra~and \suzaku~spectra.  To confirm that the larger extraction region of \suzaku~was not impacting best fitting temperatures, \chandra~and \suzaku~spectra were also fitted individually.  All values of \tx~(including the core-excised value) are consistent with one another within 1$\sigma$ errors.

\subsection{Scaled quantities}

Table~\ref{table5} lists total gravitating masses ($\rm{M}_{\rm{tot}}$), gas masses (M$_{\rm g}$), gas mass fractions (\fg), and temperatures (\tx)~within \rt. Total gravitating masses estimated from the scaling relationships M-\tx~and M-\yx~were found to be in statistical agreement with one another, with a slight tendency for the latter to be smaller.

Bolometric (unabsorbed 0.01-100 keV) luminosities \lx[\rt] were calculated by extrapolating the best fit spectrum within XSPEC.  Error bars on $\rm{L}_{\rm{X}}$ take into account uncertainties in $\rm{R}_{2500}$, temperature, abundance, and spectral normalization.  To investigate whether different methods of determining R$_{2500}$ have a significant effect on resulting \lx-\tx~relationships, we also calculate R$_{2500}$ for the \chandra~clusters using M-\tx~\citep{arnaud05}, and re-extract spectra at those radii to obtain additional luminosities.  We find that even when R$_{2500}$ values differ by 20\%, measured luminosities do not change appreciably (Table~\ref{table5}).  Since we do not expect \tx~to change drastically on $\sim50$ kpc scales at radii $R \geq$\rt~, we find that \lx-\tx~relationships are robust to different methods of determining \rt.

\section{Cluster Sample Comparisons\label{s:comp}}

\subsection{Context and comparison samples}

In our previous work \citep{hicks08} we found that high-redshift, optically selected clusters had noticeably different X-ray properties than our moderate-redshift, X-ray selected comparison sample (CNOC). With our current moderate-$z$ optically selected sample we are in a position to confirm whether these differences are due primarily to selection bias or to redshift evolution in cluster properties. We use the following comparison samples:

\begin{itemize} 

\item Ten high-redshift ($0.6 < z < 1.0$) optically selected RCS clusters selected from among the optically-richest of the 6,483 candidates detected in the first 90 deg$^2$ of the RCS \citep{gladders05}. The \chandra\ X-ray observations of these clusters are presented in \citet{hicks08}. 

\item Fourteen moderate-redshift ($0.2 < z < 0.55$) X-ray selected CNOC clusters. This sample \citep{yee96b} is derived from detections in the {\it Einstein Observatory} Extended Medium-Sensitivity Survey \citep[EMSS][]{gioia90}. CNOC was originally chosen as a comparison sample because it has been extensively observed in optical, with galaxy redshifts of $\sim1200$ cluster members as well as detailed photometric catalogues available~\citep[e.g.,][]{yee96}.  It is not, however, necessarily representative of X-ray selected samples as a whole.  As the highest-luminosity clusters from the X-ray flux-limited, wide-area EMSS survey, the CNOC sample is almost certainly drawn from the extreme X-ray bright tail of the cluster distribution. The \chandra\ X-ray observations of these clusters are detailed in \citet{hicks06}.

\item Thirty-one clusters from the \rexcess~sample \citep{bohringer07}. 
\rexcess~provides homogenous coverage of the luminosity range $0.4-20 \times10^{44}~{\rm{h}_{50}^{-2}}$ erg s$^{-1}$ in the 0.1-2.4 keV band (\tx$\geq2$ keV) over the redshift range $0.055<z<0.183$. It is representative of the X-ray selected population, clusters having been selected in X-ray luminosity only, without regard for any other characteristic (apart from angular size, so that they would fit into the field of view of {\it XMM-Newton}). The X-ray scaling properties of the \rexcess~sample within R$_{500}$ are discussed in \citet{pratt09}. For the present paper, we have recalculated the X-ray temperatures and bolometric luminosities within \rt\ defined as in Sect.~\ref{sec:scaling} above. The resulting \rt, \tx, \lx, \fg, values are listed in Table~\ref{table6}. We also list the cool core (cc) and morphological classifications \citep{pratt09}. Objects are classiÞed as cool core if central density $E(z)^{-2}\ n_{e,0} > 4 \times 10^{-2}$ cm$^{-3}$. Objects are classified as morphologically disturbed if the centroid shift parameter
\footnote{The centroid shift parameter is defined as the standard deviation of the projected separations between the X-ray peak and the centroid at each radius in the [0.1-1]~\rf~region~\citep[see e.g.,][]{pratt09,bohringer10}}
$\langle w \rangle > 0.01\, R_{500}$. The remaining systems are neither cool core nor morphologically disturbed. 
\end{itemize}

\citep{bohringer10}

\begin{table*}
\centering
\begin{minipage}{130mm}
\caption{\rexcess~X-ray Properties at $\rm{R}_{2500}$\label{table6}}
\begin{tabular}{ccccccccc}
\hline
\multicolumn{2}{c}{Cluster} & 
{$z$} & 
 {${\rm{R}}_{2500}$} & 
{${\rm{T}}_X$}   &
{${\rm{L}}_X$} &
{\fg} &
\multicolumn{1}{c}{CC} & 
{Disturbed} \\
\multicolumn{2}{c}{} & 
{} &
 {[$h_{70}^{-1}$ kpc]} & 
{[keV]} & 
{[$10^{44}~{\rm{erg}}~{\rm{s}}^{-1}$]} &
 {} &
 & \\
\hline
\multicolumn{2}{l}{RXC~J0003+0203} & 0.0924 & 388.3 & $4.24^{+0.07}_{-0.07}$ &  $1.64^{+0.01}_{-0.01}$ & 0.072 &{} & {} \\ 
\multicolumn{2}{l}{RXC~J0006-3443} & 0.1147 & 469.1 & $5.7^{+0.2}_{-0.2}$ &  $3.00^{+0.04}_{-0.04}$ & 0.068& {} & $\surd$ \\ 
\multicolumn{2}{l}{RXC~J0020-2542} & 0.1410 & 462.9 & $6.3^{+0.1}_{-0.1}$ &  $5.98^{+0.03}_{-0.03}$ & 0.095 &{} & {}\\ 
\multicolumn{2}{l}{RXC~J0049-2931} & 0.1084 & 357.7 & $3.9^{+0.2}_{-0.2}$ &  $1.68^{+0.02}_{-0.02}$ & 0.082& {} & {}\\ 
\multicolumn{2}{l}{RXC~J0145-5300} & 0.1168 & 482.4 & $5.8^{+0.1}_{-0.1}$ &  $3.86^{+0.03}_{-0.03}$  & 0.074&{} & $\surd$\\ 
\multicolumn{2}{l}{RXC~J0211-4017} & 0.1008 & 303.4& $2.20^{+0.03}_{-0.03}$ &  $0.688^{+0.005}_{-0.005}$ & 0.075&{} & {}\\ 
\multicolumn{2}{l}{RXC~J0225-2928} & 0.0604 &  307.3& $2.49^{+0.09}_{-0.01}$ &  $0.417^{+0.004}_{-0.004}$ & 0.055& {} & $\surd$\\ 
\multicolumn{2}{l}{RXC~J0345-4112} & 0.0603 &  304.9 & $2.36^{+0.04}_{-0.05}$ &  $0.701^{+0.005}_{-0.005}$ & 0.068&$\surd$ & {}\\ 
\multicolumn{2}{l}{RXC~J0547-3152} & 0.1483 & 502.1 & $6.7^{+0.1}_{-0.1}$ &  $7.87^{+0.04}_{-0.04}$ & 0.091& {} & {}\\ 
\multicolumn{2}{l}{RXC~J0605-3518} & 0.1392 & 463.2 & $4.58^{+0.04}_{-0.04}$ &  $8.67^{+0.04}_{-0.04}$  &0.102& $\surd$ & {}\\ 
\multicolumn{2}{l}{RXC~J0616-4748} & 0.1164 & 415.9 & $4.45^{+0.09}_{-0.09}$ &  $1.74^{+0.02}_{-0.02}$& 0.065&{} & $\surd$ \\ 
\multicolumn{2}{l}{RXC~J0645-5413} & 0.1644 &566.9  & $7.3^{+0.1}_{-0.1}$ &  $15.9^{+0.1}_{-0.1}$& 0.099& {} & {} \\ 
\multicolumn{2}{l}{RXC~J0821+0112} & 0.0822 & 334.8 & $3.3^{+0.1}_{-0.1}$ &  $0.650^{+0.008}_{-0.008}$ & 0.063& {} & {}\\ 
\multicolumn{2}{l}{RXC~J0958-1103} & 0.1669 &477.1 &  $5.6^{+0.3}_{-0.3}$ &  $11.0^{+0.1}_{-0.1}$ &0.107&  $\surd$ & {}\\ 
\multicolumn{2}{l}{RXC~J1044-0704} & 0.1342 & 412.7 & $3.49^{+0.02}_{-0.02}$ &  $6.91^{+0.02}_{-0.02}$ &0.112& $\surd$ & {}\\ 
\multicolumn{2}{l}{RXC~J1141-1216} & 0.1195 & 392.0 & $3.44^{+0.03}_{-0.03}$ &  $3.36^{+0.01}_{-0.01}$ & 0.087&$\surd$ & {}\\ 
\multicolumn{2}{l}{RXC~J1236-3354} & 0.0796 &  333.7& $2.82^{+0.03}_{-0.03}$ &  $0.866^{+0.006}_{-0.006}$  & 0.068&{} & {}\\ 
\multicolumn{2}{l}{RXC~J1302-0230} & 0.0847 &  372.9& $3.43^{+0.04}_{-0.04}$ &  $1.107^{+0.007}_{-0.007}$ & 0.060&$\surd$ & $\surd$ \\ 
\multicolumn{2}{l}{RXC~J1311-0120} & 0.1832 & 584.2 & $9.17^{+0.07}_{-0.07}$ &  $33.78^{+0.07}_{-0.07}$  & 0.116&$\surd$ & {}\\ 
\multicolumn{2}{l}{RXC~J1516+0005} & 0.1181 &438.4  & $5.21^{+0.07}_{-0.07}$ &  $3.59^{+0.02}_{-0.02}$ & 0.083&{} & {}\\ 
\multicolumn{2}{l}{RXC~J1516-0056} & 0.1198 & 410.5 & $4.23^{+0.09}_{-0.09}$ &  $1.60^{+0.01}_{-0.01}$  & 0.062&{} & $\surd$\\ 
\multicolumn{2}{l}{RXC~J2014-2430} & 0.1538 & 511.6 & $4.80^{+0.04}_{-0.04}$ &  $19.33^{+0.06}_{-0.06}$   & 0.110&$\surd$ & {}\\ 
\multicolumn{2}{l}{RXC~J2023-2056} & 0.0564 & 327.5 & $3.24^{+0.08}_{-0.08}$ &  $0.528^{+0.006}_{-0.006}$& 0.059&{} & $\surd$ \\ 
\multicolumn{2}{l}{RXC~J2048-1750} & 0.1475 & 477.4 & $5.3^{+0.1}_{-0.1}$ &  $3.63^{+0.03}_{-0.03}$ & 0.074&{} & $\surd$\\ 
\multicolumn{2}{l}{RXC~J2129-5048} & 0.0796 & 398.8 & $4.2^{+0.1}_{-0.1}$ &  $1.02^{+0.01}_{-0.01}$  & 0.056&{} & $\surd$\\ 
\multicolumn{2}{l}{RXC~J2149-3041} & 0.1184 & 392.6 & $3.31^{+0.03}_{-0.03}$ &  $3.10^{+0.01}_{-0.01}$ & 0.083&$\surd$ & {} \\ 
\multicolumn{2}{l}{RXC~J2157-0747} & 0.0579 & 332.8& $3.27^{+0.08}_{-0.08}$ &  $0.294^{+0.003}_{-0.003}$  & 0.041&{} & $\surd$\\ 
\multicolumn{2}{l}{RXC~J2217-3543} & 0.1486 & 452.9 & $5.38^{+0.08}_{-0.08}$ &  $5.31^{+0.02}_{-0.02}$ & 0.090&{} & {}  \\ 
\multicolumn{2}{l}{RXC~J2218-3853} & 0.1411 & 500.5 & $6.0^{+0.1}_{-0.1}$ &  $8.27^{+0.05}_{-0.05}$  & 0.095&{} & $\surd$\\ 
\multicolumn{2}{l}{RXC~J2234-3744} & 0.1510 &  568.3& $8.6^{+0.1}_{-0.1}$ &  $17.03^{+0.08}_{-0.08}$ & 0.108&{} & {} \\ 
\multicolumn{2}{l}{RXC~J2319-7313} & 0.0984 &  349.3 & $2.20^{+0.03}_{-0.03}$ &  $1.69^{+0.01}_{-0.01}$  & 0.080&$\surd$ & $\surd$\\ 
\hline

\end{tabular}
\end{minipage}
\end{table*}

\subsection{Density Profiles\label{dens}}

\begin{figure*}
\begin{center}
\includegraphics[width=0.33\textwidth]{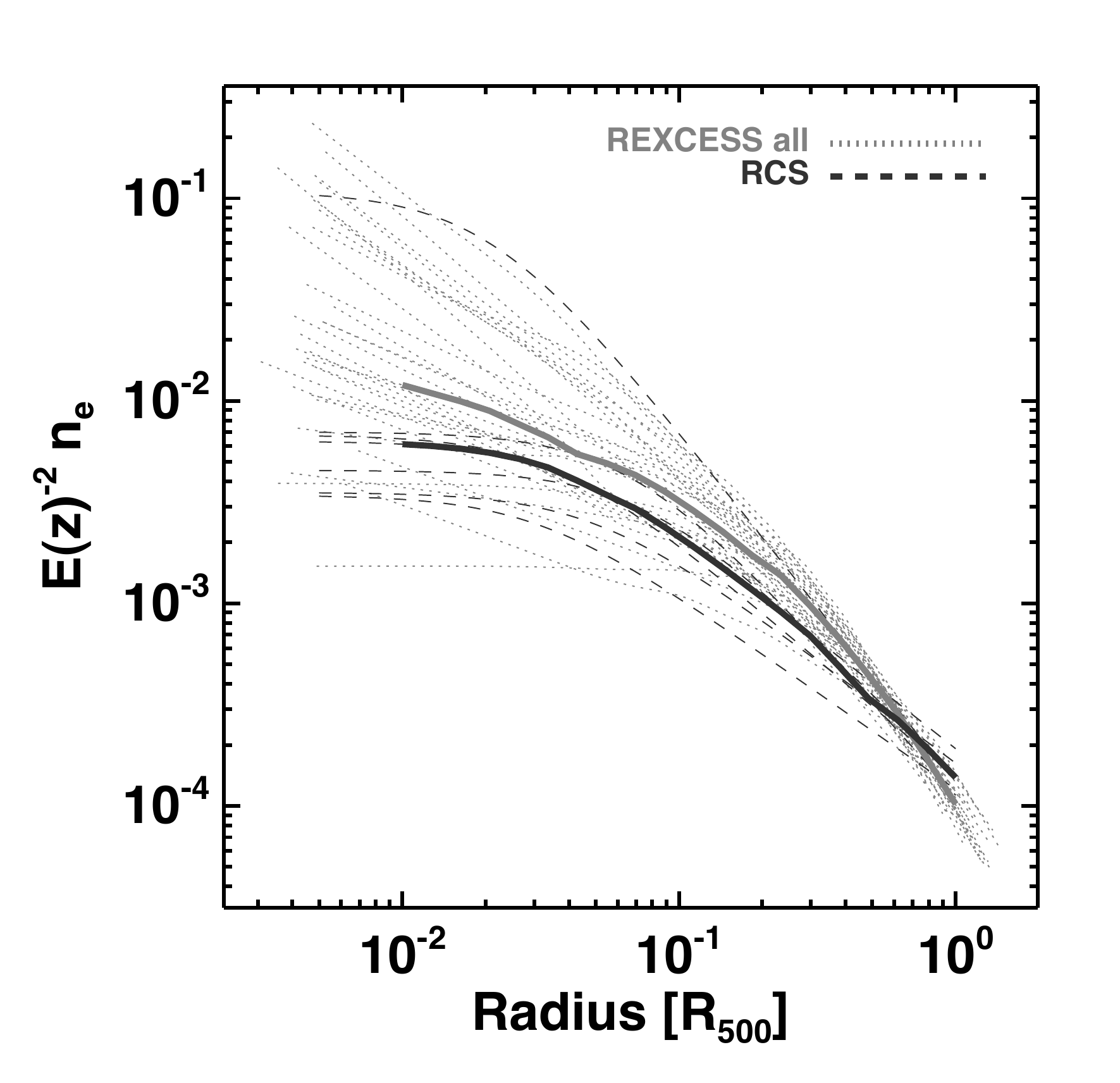}
\hfill
\includegraphics[width=0.33\textwidth]{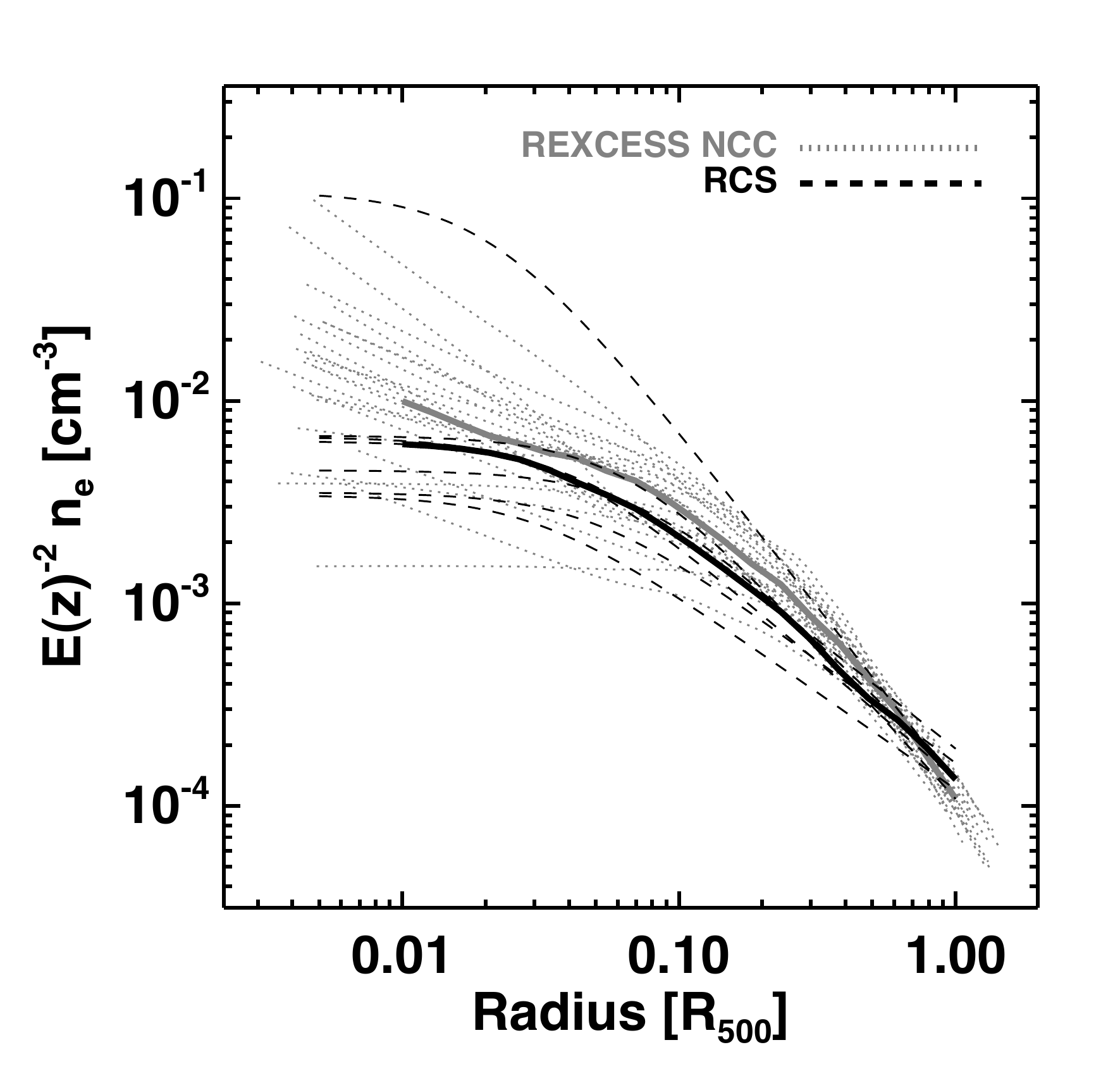}
\hfill
\includegraphics[width=0.33\textwidth]{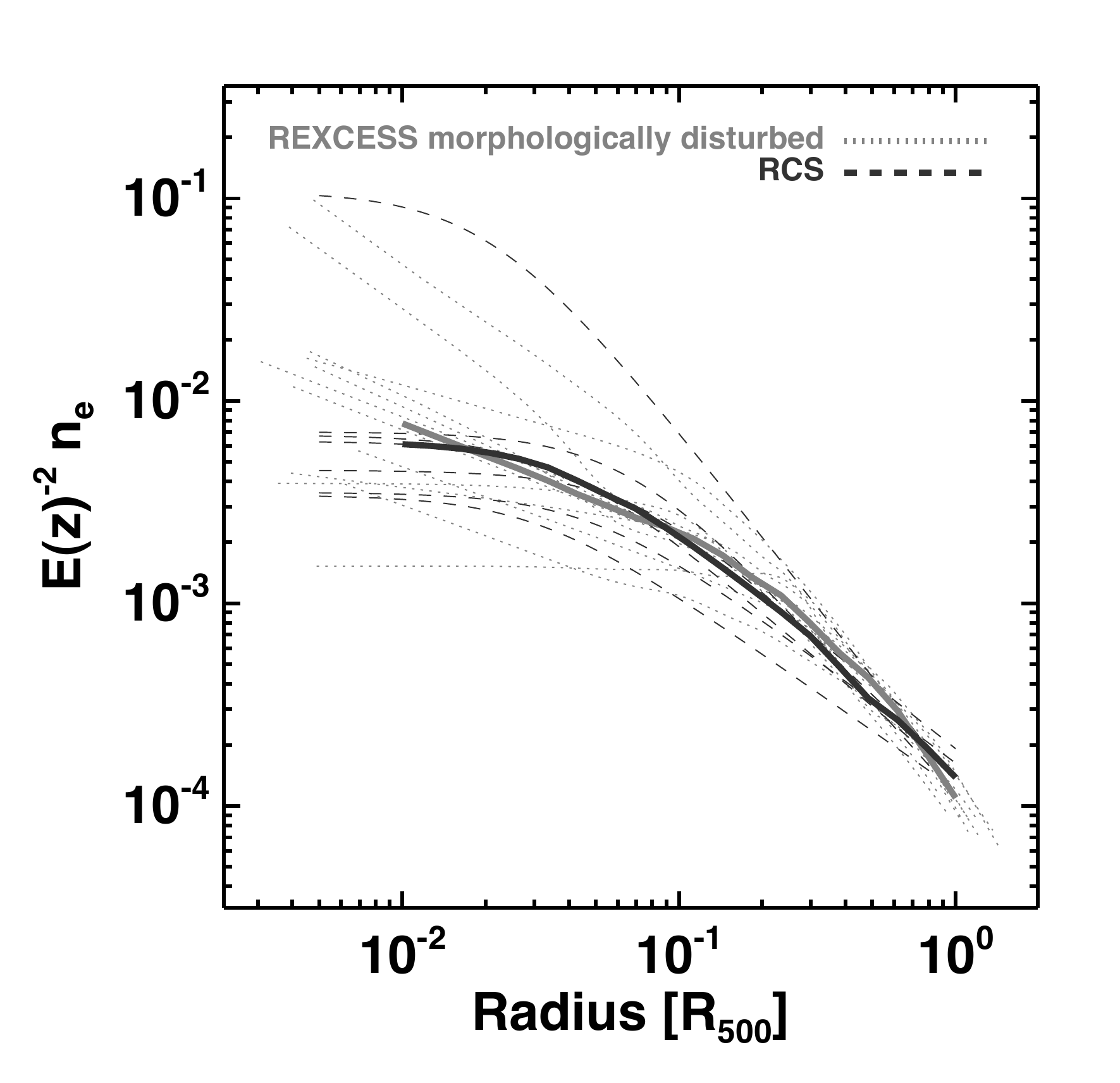}

\caption{ {\bf Density Profiles}  Moderate redshift RCS (dashed) vs \rexcess~(dotted) density profiles, including mean density profiles for each sample (solid lines).  {\it Figure 2a:} RCS vs full \rexcess~sample. The density profiles of the RCS clusters are on average flatter in their central regions than the \rexcess~systems, but are in good agreement with the X-ray systems at large radii ($\sim$\rf). {\it Figure 2b:}  RCS vs non-cool-core \rexcess~systems only, also exhibiting a slight offset. {\it Figure 2c:}  RCS vs \rexcess~morphologically disturbed subsample only. Here the mean profiles of the two samples are essentially identical.
\label{fig2}}
\end{center}
\end{figure*}

Using the $\beta$-model fits given in Table~\ref{table3} and central densities found in Table~\ref{table4}, radial density profiles were produced, scaled by~\rf, and compared to those of the \rexcess~sample. The left hand panel of Figure~\ref{fig2} shows a comparison between the scaled density profiles of the moderate-redshift RCS and \rexcess~samples. On average the scaled profiles of the RCS clusters are suppressed in the central regions compared to \rexcess. Note that this is a strong effect, reaching out well beyond \rt.  RCS1447 is included in the calculation of the mean profile. Excluding it from the calculation would further suppress the average scaled density profile of the RCS systems in the central regions.

The middle panel of Figure~\ref{fig2} shows the RCS clusters compared to the non-cool-core \rexcess~subset only. Once again, there is a clear offset in the average density profile that extends out well beyond \rt.

The right hand panel of Fig. 4 shows the RCS clusters compared to the morphologically disturbed \rexcess~subsample. Here the agreement between mean scaled density profiles is actually quite remarkable. Note that there are two cool-core systems in the \rexcess~morphologically disturbed subset, just as there is a cool core in the RCS moderate-z sample (RCS1447), thus both samples span the whole range of cluster properties. Also, the mass ranges of the two samples match quite well, and this good agreement in mass is reflected in the similar temperature range. Therefore this effect is not likely to be the result of variations in gas mass fraction with mass.

\subsection{The $L_X-T_X$ Relationship\label{s:lxtx}}

Studying the relationships between global cluster properties (L$_X$, T$_X$, M$_{\rm{tot}}$, etc.) over a broad range in redshift allows us to investigate the influence of non-gravitational processes on cluster formation and evolution. In addition, these relationships give us insight regarding cluster dynamical state and composition, as well as provide ready methods of comparison between different cluster samples.  In this work we focus primarily on X-ray temperature (\tx) and luminosity (\lx), since they are tied most closely to the actual data and require fewer assumptions than extrapolated properties such as total cluster mass.  To facilitate comparisons between optically selected and X-ray selected clusters, we make use of our previous {\it{Chandra}} analyses of the CNOC~\citep{hicks06} and high-$z$ RCS samples~\citep{hicks08}.  

In all \lx-\tx~relationships, \lx~has been scaled by the cosmological factor $E_z={H(z)}/{H_0} = \left[{\Omega_m (1+z)^3 + \Omega_\Lambda} \right]^{1/2}$ and the relationship has been fitted with the form

\begin{equation}
{\rm{log}_{10}}~{(E_z^{-1}\rm{L_X})}=C1+C2~{\rm{log}_{10}}~{\rm{T_X}},
\end{equation}

\noindent  with \tx~in units of 5 keV and \lx~in units of $10^{44}$ erg s$^{-1}$.  
New best fitting relationships are determined using both the bisector and orthogonal modifications of the BCES algorithm in~\citet{akritas}, with figures displaying bisector fits for ease of comparison to our previous work.

\begin{figure*}
\centerline{\includegraphics[width=3in, angle=90]{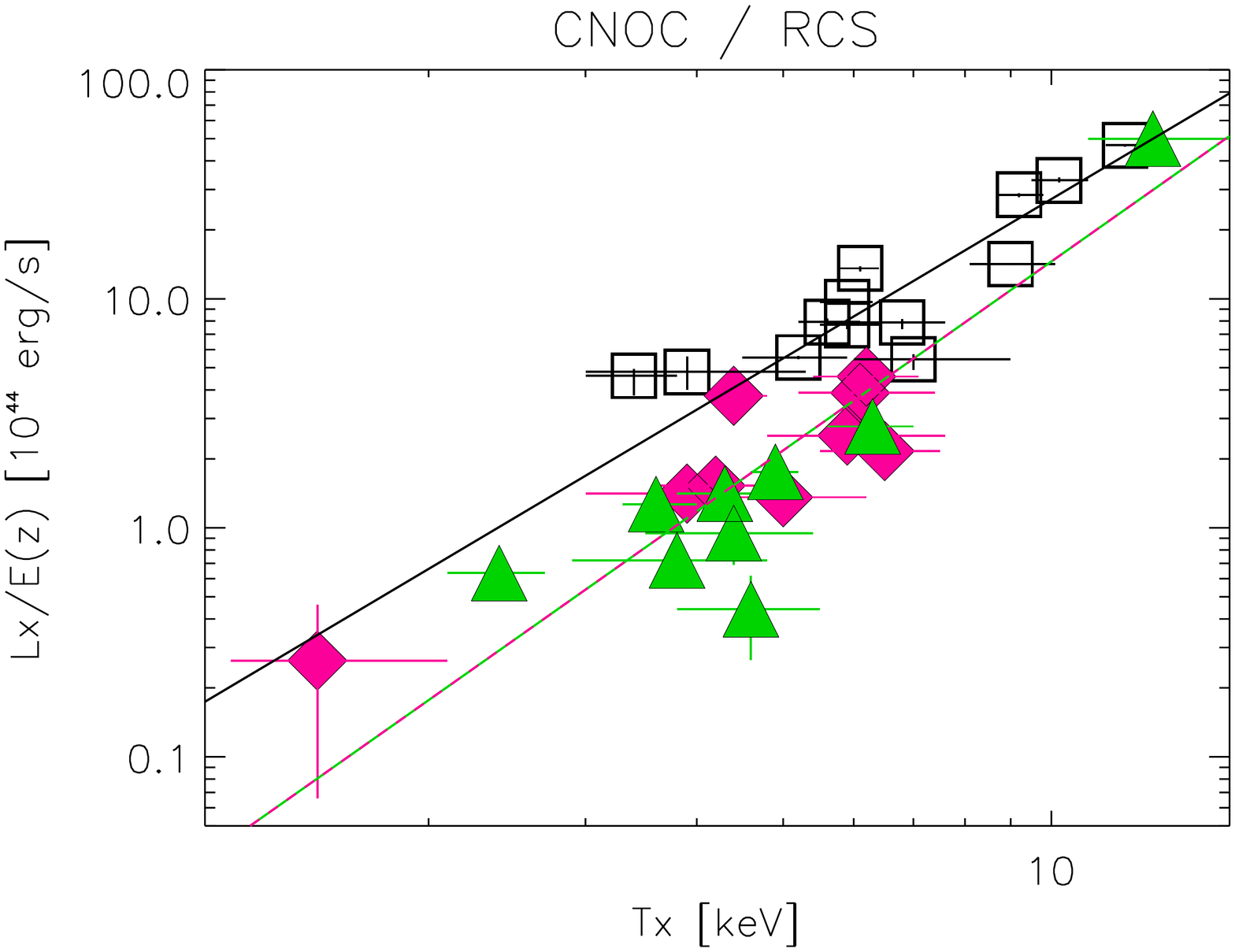}
\includegraphics[width=3in, angle=90]{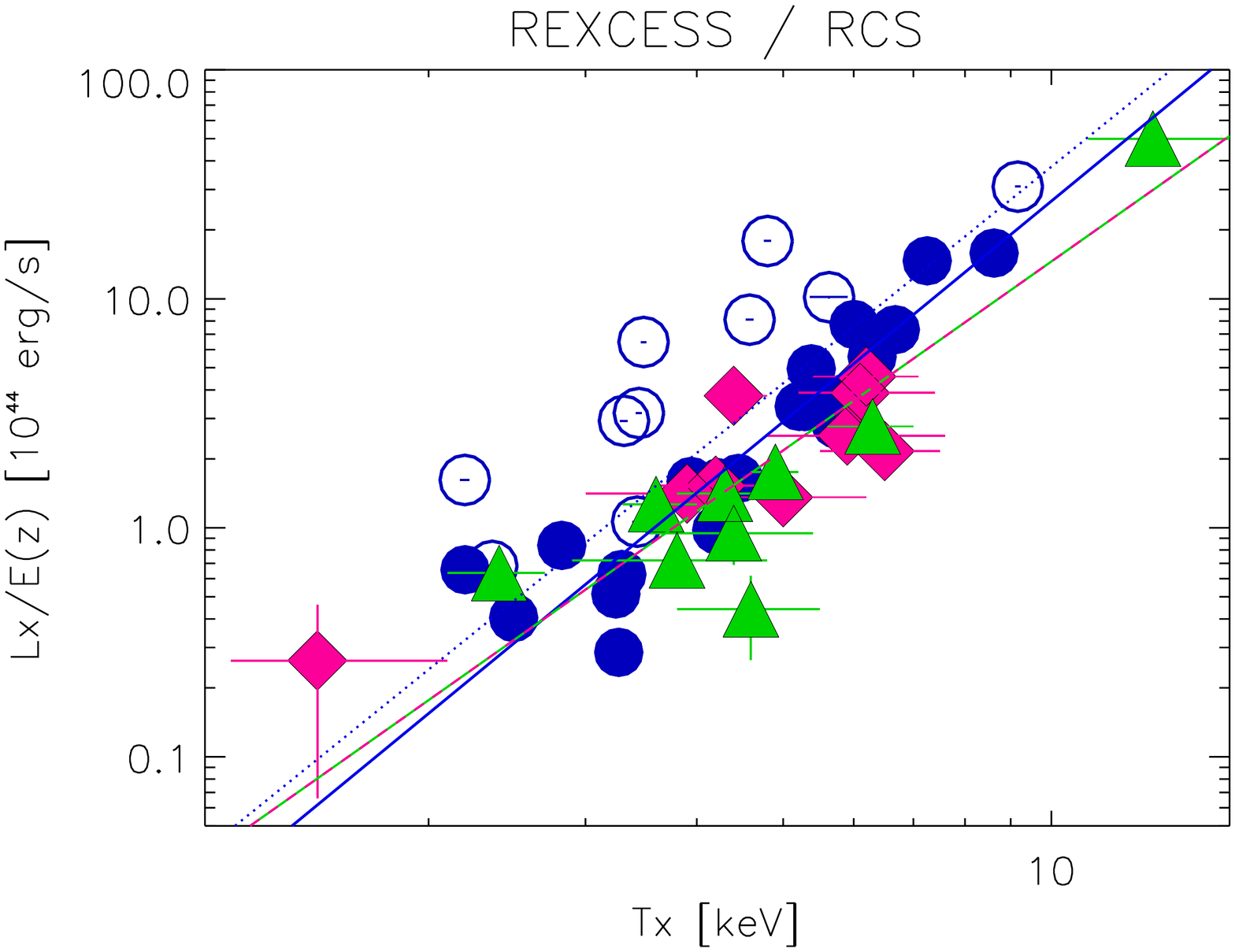}}
\centerline{\includegraphics[width=3in, angle=90]{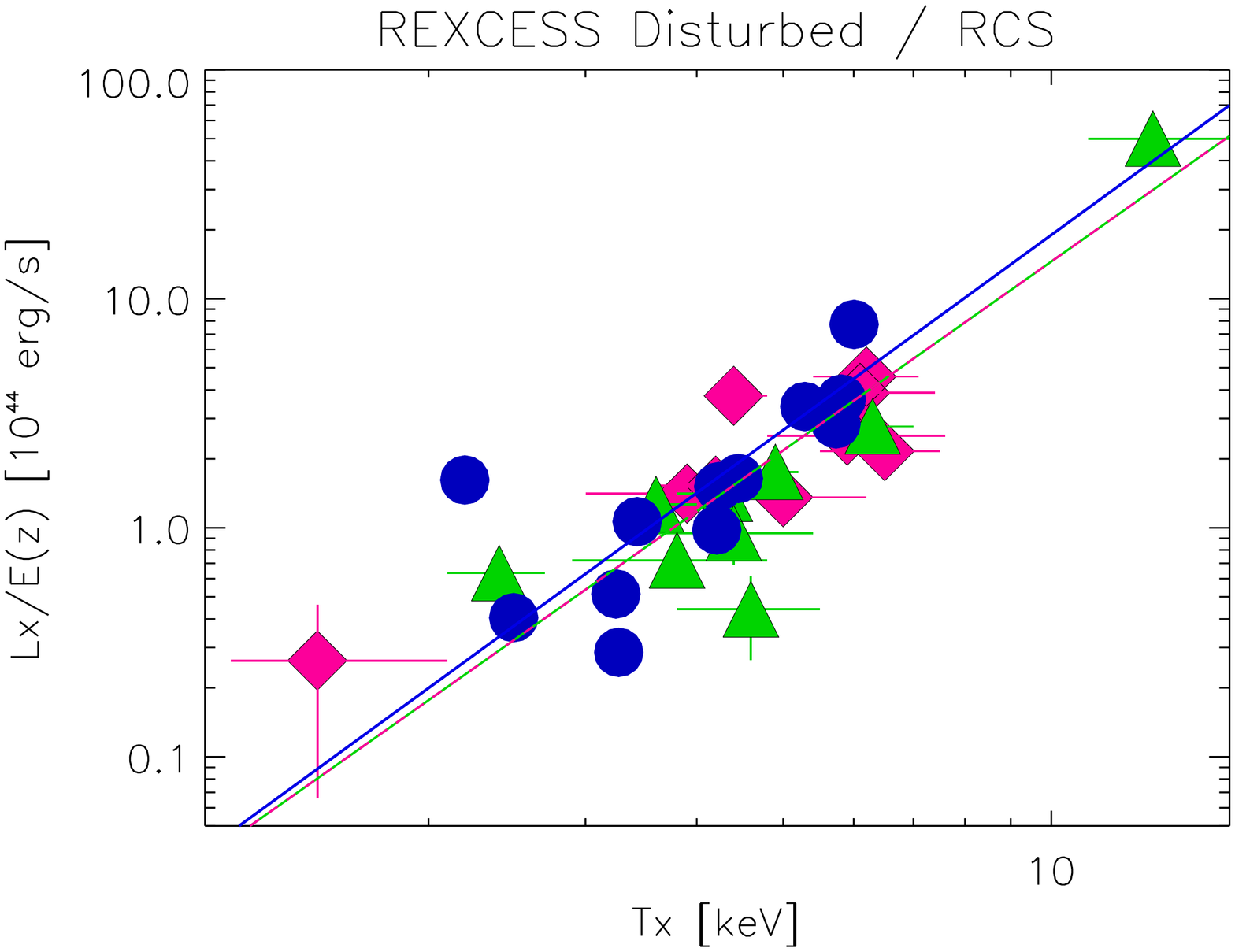}}
\caption{{\bf{\lx-\tx~Relationships.}}  X-ray temperatures are plotted against cosmologically consistent, intrinsic bolometric luminosities within $\rm{R}_{2500}$.  In all panels diamonds represent high-$z$ RCS clusters ($z_{\rm{avg}}=0.80$), triangles denote the current sample ($z_{\rm{avg}}=0.30$), and the dashed line shows the \lx-\tx~relationship for the entire X-ray observed RCS sample.
{\it{Figure 3a:}}  Squares designate moderate redshift CNOC clusters ($z_{\rm{avg}}=0.32$), and the solid line shows their best fitting relationship.  As in~\citet{hicks08}, the X-ray selected sample exhibits a higher normalization than optically selected clusters.  {\it{Figure 3b:}}  Circles represent the \rexcess~sample.  Here the solid line shows the fit to non-cool-core \rexcess~clusters (filled circles) and the dotted line is the best fit to the entire \rexcess~sample.  Open circles denote cool-core clusters.  {\it{Figure 3c:}}  Here circles show only the disturbed \rexcess~clusters, and the solid line represents their best fitting \lx-\tx~relationship, which is notably similar to that of the RCS combined sample (dashed line). 
\label{fig3}}
\end{figure*}

It is immediately clear from visual inspection of Figure~\ref{fig3}a that our current targets are overall less luminous for a given temperature than the CNOC sample, in keeping with our previously-observed RCS sample.  This result suggests that global discrepancies between the X-ray properties of X-ray and optically selected cluster samples are wholly attributable to selection bias.

After correcting for self-similar evolution we find no convincing evidence of redshift evolution in cluster properties within the complete X-ray observed RCS sample ($0.15<z<1.0$).  To further explore this point, we fitted the RCS moderate-$z$ and high-$z$ samples together with the function

\begin{equation}
{\rm{log}_{10}}~({\rm{L_X}}) = A~{\rm{log}_{10}}~(E_z/1.3) + B~{\rm{log}_{10}}~({\rm{T_X}}/5) + C
\end{equation}

The best fitting values are (bootstrap uncertainties): $A = 0.43\pm{1.15}$; $B = 2.69\pm{0.84}$; and $C = 3.23\pm{1.24}$, showing that we are quantitatively unable to constrain evolution with the current sample.

Figure~\ref{fig3}b shows RCS and \rexcess~\lx-\tx~scaling relationships, for the entire \rexcess~sample and for just the non-cool-core (ncc) clusters.  For the \rexcess~sample, cool cores define the upper envelope of the relation while morphologically disturbed systems define the lower envelope.  The vast majority of the dispersion comes from the cool cores. While \rexcess~cool-core clusters look much like the high-\lx~CNOC sample, the non-cool-core clusters are situated similarly to the optically selected sample.  

It is clear that the normalization, in particular, that one will obtain for any \lx-\tx~fit will depend strongly on the sample composition.  Note also that the intrinsic dispersion about the \rexcess~\lx-\tx~relation is considerably larger than the 70\% dispersion about the \lx-\tx~relation for quantities measured within \rf~\citep{pratt09}. The larger dispersion we see is due to measuring quantities in the core regions, where the variation in density is strongest.

Building on the results of Section~5.2, we compare \rexcess~morphologically disturbed clusters to the RCS sample (Figure~\ref{fig3}c).  Here we find an even closer similarity in X-ray properties, to the extent that even best-fitting relationships lie nearly on top of one another.  This result is expected given the agreement in gas density profiles between our targets and the \rexcess~morphologically disturbed subsample.  \lx-\tx~fits for a variety of sample combinations are reported in Table~\ref{table7}.

\begin{table*}
\centering
\begin{minipage}{100mm}
\caption{Scaling Relationship: $\rm{log}_{10}~{(E_z^{-1}\rm{L_X})}=C1+C2~{\rm{log}_{10}}~{\rm{T_X}}$\label{table7}}
\begin{tabular}{ccccc}             
\hline   
{Sample} & \multicolumn{2}{c} {\underline{Bisector}} &\multicolumn{2}{c}{\underline{Orthogonal}} \\   
{} &
 {$C_1$} & 
 {$C_2$} &
 {$C_1$} & 
 {$C_2$} \\
\hline
    \multicolumn{1}{l}  {CNOC\footnote[1]{\citet{hicks08}}} & $0.74\pm{0.08}$ & $2.3\pm{0.3}$& {} & {} \\    
     \multicolumn{1}{l}{RCS (high-$z$)$^{\displaystyle a}$} & $0.36\pm{0.06}$ & $2.1\pm{0.3}$& {} & {} \\
    \multicolumn{1}{l}{RCS (total)} & $0.34\pm{0.06}$ & $2.7\pm{0.5}$& $0.34\pm{0.06}$ &$2.8\pm{0.5}$ \\
       \multicolumn{1}{l}  {\rexcess~(disturbed)} & $0.43\pm{0.05}$ & $2.8\pm{0.7}$& $0.5\pm{0.1}$ & $3.8\pm{0.6}$\\
         \multicolumn{1}{l}{\rexcess~(ncc)} & $0.46\pm{0.03}$ & $3.2\pm{0.3}$& $0.47\pm{0.03}$ & $3.4\pm{0.4}$\\
         \multicolumn{1}{l}{\rexcess~(total)} & $0.63\pm{0.06}$ & $3.1\pm{0.3}$& $0.68\pm{0.07}$ & $3.9\pm{0.5}$ \\
\hline
\end{tabular}
\end{minipage}
\end{table*}

\subsection{Gas Mass Fractions\label{fgas}}

Using our estimated masses (Table~\ref{table5}) we calculate core gas mass fractions within $\rm{R}_{2500}$.  Our sample mean is $.058\pm{0.008}$ using \yx~to determine total mass (error computed by dividing the standard deviation by $\sqrt{N}$).  In comparison, \rexcess~clusters have a mean gas mass fraction of $0.081\pm{0.003}$ for the complete sample of 31 clusters, and $0.066\pm{0.004}$ when including only the 12 disturbed clusters; further evidence of similarities between optically selected clusters and morphologically disturbed X-ray selected clusters. 

\begin{figure*}
\centerline{\includegraphics[width=3.5in, angle=0]{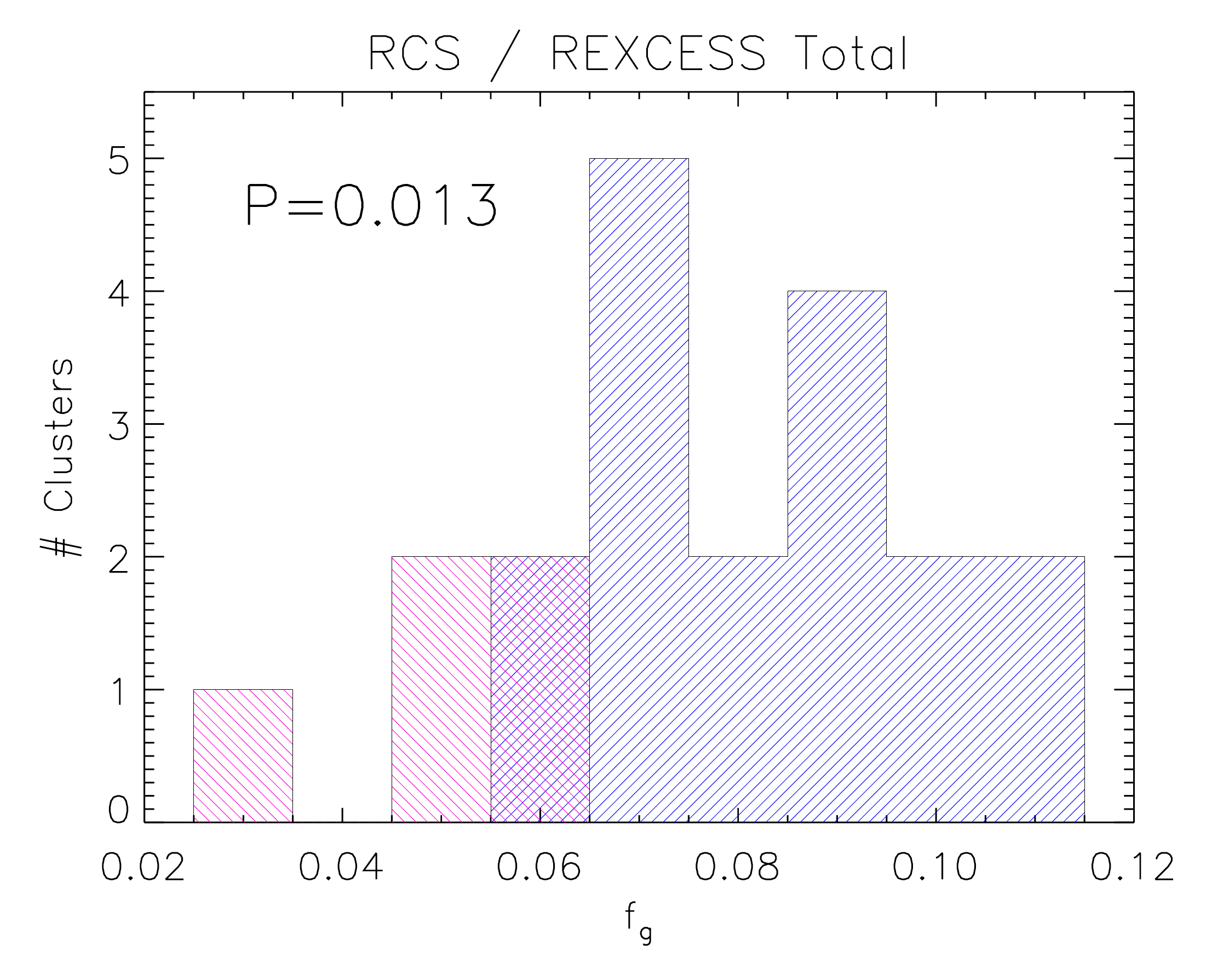}
\includegraphics[width=3.5in, angle=0]{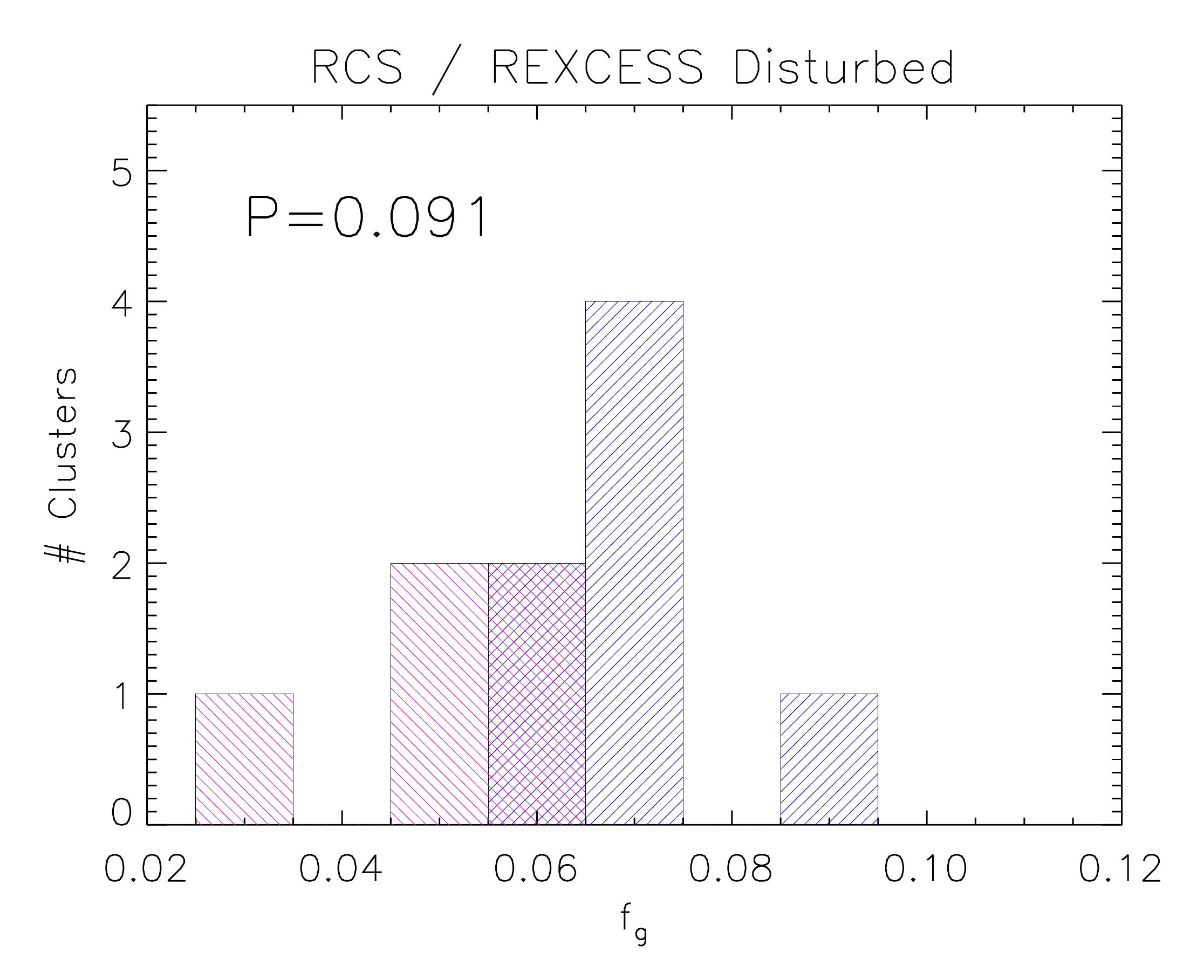}}
\caption{{\bf{Gas Mass Fractions.}}
{\it{Left:}} Histogram of gas mass fractions for the 5 moderate-$z$ RCS and 17 \rexcess~clusters with $3.5<$\tx$<8$ keV. A K-S test performed on these two samples resulted in D=0.741 and P=0.013, indicating that the gas mass fractions of the two samples are unlikely to have been drawn from the same distribution.  {\it{Right:}} Using only the 7 disturbed \rexcess~clusters with $3.5<$\tx$<8$ keV.
 \label{fig4}}
\end{figure*}

We perform statistical tests comparing the distributions of \fg~values for a subset of the clusters in our moderate redshift RCS sample with two subsets of the \rexcess~sample.  In order to compare clusters of similar masses, and to be consistent with~\citet{hicks08}, we restricted the temperature range in all 3 samples to be between 3.5-8.0 keV, resulting in 5 clusters from the RCS, 17 clusters in the \rexcess~sample, and 7 clusters in the \rexcess/disturbed sample.
The mean gas fraction and standard deviation from each of the subsamples are $0.050\pm{0.005}$, $0.084\pm{0.004}$, and $0.071\pm{0.005}$, for the RCS subset, \rexcess~subset, and \rexcess/disturbed subset respectively. The RCS subset
shows lower \fg~than the \rexcess~cluster subset, but similar \fg~to the \rexcess/disturbed cluster subset, also visible in the histograms shown in Figure~\ref{fig4}.

We also performed a two-sided Kolmogorov-Smirnov test~\citep[the contributed KSTWO IDL routine, based on][ Numerical Recipes]{press92}, pairing the RCS subset with each of the \rexcess~and  \rexcess/disturbed subsets. The KS test comparing RCS clusters with the complete \rexcess~sample showed a higher probability of having been randomly drawn from different parent populations than the test involving only the \rexcess/disturbed subsample; however, such a comparison is severely limited by the small number of clusters.  The KS probability comparing the 5 RCS clusters with the 17 \rexcess~clusters was 0.013, while the KS probability comparing the RCS clusters with 7 \rexcess/disturbed clusters was 0.091. 

\section{Summary and Discussion \label{s:sum}}

We have presented a detailed X-ray investigation of a sample of moderate-redshift, optically selected clusters of galaxies from the Red-sequence Cluster Survey (Table~\ref{table1}).  All of our targets were detected by \chandra~and/or \suzaku~at S/N$>$11 (Table~\ref{table2}) and were found within 27\arcsec~of their optical positions.  \chandra~imaging reveals that most of these clusters possess at least some degree of substructure~(Figure~\ref{fig1}).
 
Surface brightness profiles were extracted for the seven \chandra-observed clusters in S/N$>$3 annular bins.  These profiles were reasonably well fitted by single $\beta$-models (Table~\ref{table3} and Figure~\ref{figB1}).  Cluster emission was modeled with XSPEC, beginning with a spectral extraction region of 300 kpc radius. The results of single-temperature spectral fits (Table~\ref{table4}), combined with gas masses obtained using best-fit $\beta$-models, were used to determine \yx~and consequently \rt.  This process was carried out iteratively until extraction regions and \rt~estimates were in agreement. 

Using both the M-\yx~\citep{arnaud07} and M-\tx~relationship~\citep{arnaud05}, total masses were estimated out to \rt.  While both relationships are overall consistent within our errors, we find M-\yx~mass estimates to generally be lower than those obtained via M-\tx~for the clusters in our sample, not entirely surprising given that M-\yx~uses gas mass information as well as temperature.  Reassuringly, uncertainties in \rt~do not lead to large uncertainties in \lx[\rt]. Therefore for X-ray observations for which only global \tx~and \lx~measurements are possible due to poor spatial resolution and/or low net counts, using M-\tx~to determine~\rt~results in reasonably accurate \lx-\tx~relationships. Values of \rt, total mass, gas mass, gas mass fraction, and integrated luminosity are given in Table~\ref{table5}. 
 
The scaled density profiles (Figure~\ref{fig2}) and global quantities of the moderate-redshift RCS sample are most consistent with the morphologically disturbed subset of the \rexcess~X-ray selected sample. These density profiles explain trends in \lx-\tx relationships (e.g., Figure~\ref{fig3};~\citealt{hicks08}) without having to resort to non-standard evolutionary effects. This result suggests that much of the non-standard evolution found by previous work may be due to the varying fractions of cool cores in different samples, which in turn are affected by the various methods of cluster selection. 

The properties of the scaled density profiles are reflected in the global quantities \tx~and especially \lx, and vice versa. Low \lx~systems have centrally-suppressed density profiles; high \lx~systems have centrally-peaked density profiles. Trends in the \lx-\tx~relation measured in an aperture corresponding to \rt~are similar to those measured at \rf~\citep{pratt09}. cool-core systems populate the high-\lx~side of the relation; morphologically disturbed systems populate the low-\lx~side.

The fraction of cool-core clusters in the low-redshift RCS sample appears to be lower than in the \rexcess~sample (1/7 versus 10/31). This lower fraction is consistent with the differences seen in density profiles and scaling relations. However, the small numbers and potential mismatch in mass distributions keep this discrepancy in cool-core fraction from being statistically significant. If there is an increase in the cool-core fraction towards the present day, there is a possibility that the percentage of cool cores found in X-ray and optically selected samples might converge at higher redshift. However, there are very few minimally-biased X-ray selected cluster samples that probe the redshift ranges we are interested in. The eROSITA survey \citep{predehl11} will definitely shed light on this issue. 
  
Overall we find that optically selected RCS cluster properties span the entire range of those of massive clusters selected via other methods, but contain a higher fraction of objects with gas density profiles and global properties similar to those of the morphologically-disturbed systems in \rexcess. This result suggests that optical and X-ray selection do not sample exactly the same population of clusters. Recent results from SZ surveys such as Planck  \citep{planck11b}, further suggest that X-ray selection may preferentially pick up centrally-concentrated systems. Selection effects such as these have the potential to affect the interpretation of  cluster surveys that intend to use the evolution of the cluster population as a proxy for cosmic evolution. More investigations are clearly necessary.

\section*{acknowledgements} 

Support for this work was provided by the National Aeronautics and Space Administration through \suzaku~Awards NNX08AZ72G and NNX10AH92G, and \chandra~award GO9-0149 issued by the \chandra~X-ray Observatory Center, which is operated by the Smithsonian Astrophysical Observatory for and on behalf of the National Aeronautics Space Administration under contract NAS8-03060. NSF AST-0206154 provided support for spectroscopic observations.  Support for Megan Donahue and Amalia Hicks was also provided by LTSA award NNG05GD82G. 










\appendix

\section{Signal-to-Noise Ratios}\label{s:sig}

To estimate the significance of RCS cluster detections in the X-ray, we used the same simple statistics as in~\citet{hicks08}.  
For the \chandra~observed clusters, counts were extracted in the 0.3-7.0 keV band from a 500 $h_{70}^{-1}$ kpc radius region around the X-ray peak ($C$) and also from a region away from the aimpoint on the same chip which served as a background ($B$).  Due to the $\sim2$ arcminute spatial resolution of \suzaku, all analysis was performed using a 260 arcsecond extraction radius, as recommended by the \suzaku~Data Reduction Guide~\footnote{http://heasarc.nasa.gov/docs/suzaku/analysis/abc/}.  Therefore to determine S/N for our \suzaku~observations, counts were extracted in the 0.2-12.0 keV band from a 260\arcsec~radius region centered on the cluster ($C$), and the remainder of the chip minus calibration sources was used as the background ($B$).  Obvious point sources were removed from each region in all observations.  Final signal-to-noise ratios were calculated based on dividing net counts, $N=C-B$, by the standard deviation, $\sigma=\sqrt{C+B}$.  Using this method, all clusters were solidly detected at a signal-to-noise ratio greater than 11 (Table~\ref{table2}), with an average S/N of 41 for the sample.

\section{$\beta$ model fits}

\begin{figure*}
\centering
\includegraphics[width=7in]{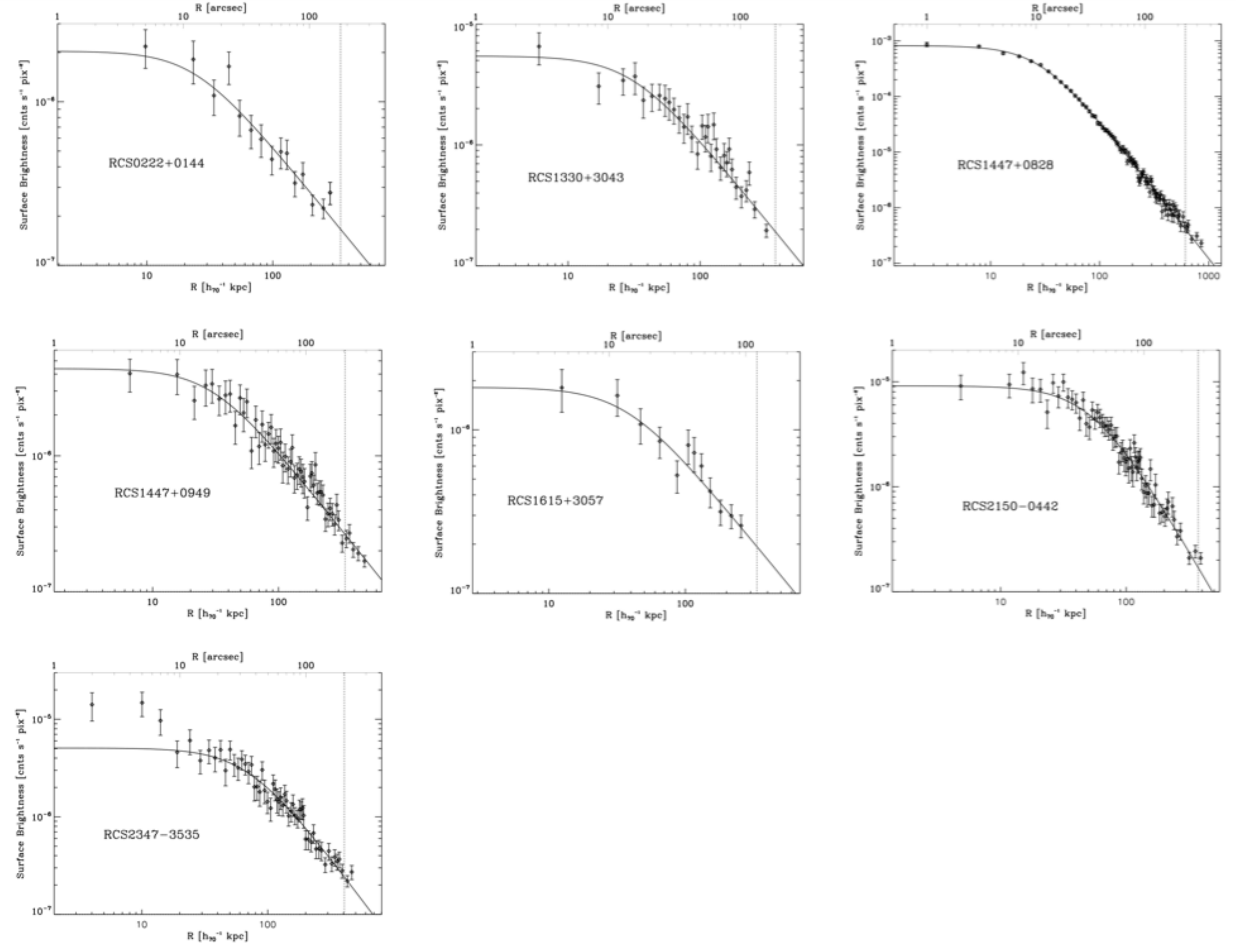}
\caption{{\bf{Surface Brightness Profiles.}}  Background-subtracted radial surface brighness profiles for the 0.3-2.5 keV band accumulated in S/N$>3$ annular bins for seven of the nine clusters in our sample.  
A solid line traces the best fitting single $\beta$ model of each cluster.  
Vertical dashed lines indicate R$_{2500}$.  
Many of the profiles exhibit some substructure; however, most were reasonably well fitted by a standard $\beta$ model (see Table~\ref{table3} for goodness of fit data).  The surface brightness profile of RCS2347-3535 suggests that it harbors a modest cool core in the inner 10-20 kpc.\label{figB1}}
\end{figure*}





\end{document}